\documentclass[11pt,a4paper,public]{redpants}
\pdfoutput=1

\usepackage{bm}
\usepackage{color}
\usepackage{graphicx}
\usepackage{array}
\usepackage{float} 
\usepackage{amsmath,amsfonts}
\usepackage{verbatim}

\newcommand{\mRe}{{\rm Re}}
\newcommand{\mIm}{{\rm Im}}

\newcommand{\rmd}{{\rm d}}

\newcommand{\myarraystretch}{2}

\newsavebox{\mycardbox}

\newenvironment{stabular*}[2]{%
\renewcommand\arraystretch{#1}%
\begin{center}
\footnotesize
\begin{lrbox}{\mycardbox}
\begin{tabular}{#2}%
}{%
\end{tabular}
\end{lrbox}\resizebox{\textwidth}{!}{\usebox{\mycardbox}}
\renewcommand\arraystretch{\oldarraystretch}%
\end{center}}

\newenvironment{stabular}[1]{%
\begin{stabular*}{\myarraystretch}{#1}%
}{%
\end{stabular*}
}

\def\imagetop#1{\vtop{\null\hbox{#1}}}

\newcommand{\lwig}{\mbox{\;\raisebox{.3ex}
    {$<$}$\!\!\!\!\!$\raisebox{-.9ex}{$\sim$}\;}}

\newcommand{\lambdabar}{{\hbox{$\lambda$\kern-1.ex\raise+0.45ex\hbox{--}}}}

\DeclareMathAlphabet{\mathpzc}{OT1}{pzc}{m}{it}

\title{How real-time cosmology can distinguish between different anisotropic models}

\author[a]{Luca Amendola,}
\emailAdd{{l.amendola@thphys.uni-heidelberg.de}}
\author[b]{Ole Eggers Bj\ae lde,}
\emailAdd{{oeb@phys.au.dk}}
\author[c]{Wessel Valkenburg,}
\emailAdd{{valkenburg@lorentz.leidenuniv.nl}}
\author[d]{and Yvonne~Y.~Y.~Wong}
\emailAdd{{yvonne.y.wong@unsw.edu.au}}

\affiliation[a]{Institut f\"ur Theoretische Physik, Universit\"at Heidelberg \\ Philosophenweg 16, D-69120 Heidelberg, Germany}
\affiliation[b]{Department of Physics and Astronomy, Aarhus University \\ Ny Munkegade 120, DK-8000 Aarhus C, Denmark}
\affiliation[c]{Instituut-Lorentz for Theoretical Physics, Universiteit Leiden \\ Niels Bohrweg 2, 2333 CA Leiden, The Netherlands}
\affiliation[d]{School of Physics, The University of New South Wales\\
Sydney NSW 2052, Australia}

\begin{document}

\maketitle
\begin{abstract}
We present a new analysis on how to distinguish between isotropic and anisotropic cosmological models  based on tracking the angular displacements of a large number of distant quasars over an extended period of time, and then performing a multipole-vector decomposition of the resulting displacement maps.
We find that while the GAIA mission operating at its nominal specifications does not have sufficient angular resolution to resolve anisotropic universes from isotropic ones using this method within a reasonable timespan of ten years, a next-generation GAIA-like survey with a resolution ten times better should be equal to the task.  Distinguishing between different anisotropic models is however more demanding. Keeping the observational timespan to ten years, we find that the angular resolution of the survey will need to be of order 0.1~$\mu$as in order for certain rotating anisotropic models to produce a detectable signature that is also unique to models of this class.  However, should such a detection
become possible, it would immediately allow us to rule out large local void models.
\end{abstract}

\section{Introduction}

In the current era of precision cosmology, the foundations of the so-called standard model of cosmology are constantly being challenged by observations of
 ever-increasing accuracy and precision. This standard model describes the universe as homogeneous and isotropic on large length scales, whose energy content consists
presently of approximately  75\% dark energy which drives the accelerated expansion of the universe, and  25\%  nonrelativistic matter.
These simple deductions are the culmination of many different types of observations, and the timescales on which the numbers evolve substantially are of order the age of the universe.

In the last years, however, a new way of inferring and testing the global properties of the universe on a much shorter timescale has emerged: real-time cosmology
(RTC)~\cite{Quercellini:2008ty,Quercellini:2010zr}.
The basic idea is to take advantage of the high level of precision offered by current measurements of  angular positions and redshifts for objects such as quasars in the universe.
Because the movement of such objects is cosmology-dependent, tracking the same objects over an extended period of time, e.g., ten years, may help shed light
on the underlying cosmology~\cite{Quartin:2009xr,Quercellini:2009ni,Fontanini:2009qq,Ding:2009xs,Campanelli:2011wz}.

The main focus of earlier works on RTC has been to distinguish between a completely isotropic universe and a universe wherein some degree of large-scale anisotropy exists.
Examples of anisotropic universes include Lema\^{i}tre--Tolman--Bondi (LTB) void models~\cite{Quartin:2009xr}, meatball universes~\cite{Park:1991wp,Shandarin:1997fc}, and Bianchi universes~\cite{Quercellini:2009ni}.  Universes in which the observer resides in a massive overdensity such as the hypothetical Great Attractor also belong to this group of models~\cite{Valkenburg:2012ds}. What has not be established so far, however,
is if RTC could differentiate these anisotropic models from one another.%
\footnote{Note that only strongly anisotropic models have an impact on RTC. Milder anisotropies, such as those arising from  models that combine dark energy with LTB~\cite{Sinclair:2010sb, Marra:2010pg, deLavallaz:2011tj, Romano:2011mx,Valkenburg:2011ty,Valkenburg:2012ds}, are not detectable.}

In this paper, we explore the potential of RTC further, and propose a new test based on multipole alignment that may help distinguish between different classes of large-scale
anisotropic universes.  The basic premise is that the movements of the observed objects can be encoded in a set of ``displacement maps'', and the maps analysed using a multipole expansion.  Because each class of anisotropic models predicts a unique set of  multipoles---especially in terms of their alignments---various classes of models can be differentiated from one another
on this basis.
We illustrate the methodology in the following using two toy anisotropic models. The first is a radially expanding spherically symmetric LTB void model, which has been investigated extensively in the literature (see, e.g., \cite{Biswas:2010xm} and references therein). The second is a universe in which the observer lives inside a large region with a net rotation, while outside this region the net rotation is zero.   Substantial global rotation is not generally expected in standard cosmologies, since any rotation will be heavily diluted in the process of cosmological inflation~\cite{Ellis:1982xw}. However, extended regions of space such as voids or giant clusters of galaxies can possess some degree of angular momentum~\cite{Hwang:2007ba}. Large-scale rotation models and models with global rotation in general have been investigated to some extent in the literature (see, e.g.,~\cite{Barrow:1985tda,Li:1997du} and references therein).

The paper is organised as follows. In section~\ref{sec:II}, we present the formalism for the construction of the displacement maps and their associated spherical harmonic moments and multipole vectors.  The formalism is then applied to LTB and rotating universes in section~\ref{sec:examples}, in order to establish their model signatures.
In section~\ref{sec:V}, we describe the method of our numerical simulations, the results of which we discuss in~\ref{sec:forecast} for a selection of models based on the expected performance of the GAIA quasar survey~\cite{Jordi:2005te}.  Section~\ref{sec:VI} contains our conclusions.  Our numerical code is available for download at
\url{http://web.physik.rwth-aachen.de/download/valkenburg/Rotation/}, and can be used for any model of object displacement.

\section{Formalism}\label{sec:II}

We begin with the expression for the angular separation between two objects as measured by an observer,
\begin{equation}
\cos \gamma_{12} = \frac{ (\vec{s}_1-\vec{s}_o) \cdot (\vec{s}_2-\vec{s}_o)}{|\vec{s}_1-\vec{s}_o||\vec{s}_2-\vec{s}_o|} \equiv \hat{n}_1 \cdot \hat{n}_2,
\end{equation}
where $\vec{s}_1$ and $\vec{s}_2$ denote the positions of the two objects with respect to some coordinate system, $\vec{s}_o$ the location of the observer, and $\vec{n}_i \equiv \vec{s}_i-\vec{s}_o$.  If each object (including the observer) moves to a different coordinate point in space, i.e.,
$\vec{s}_i \to \vec{s'}_i = \vec{s}_i + \vec{\Delta s}_i$,
then the new angular separation $\gamma_{12}'=\gamma_{12} + \Delta \gamma_{12}$ between the objects as seen by the observer will be given by
\begin{equation}
\cos \gamma_{12}' = \frac{ (\vec{s'}_1-\vec{s'_o}) \cdot (\vec{s'}_2-\vec{s'}_o)}{|\vec{s'}_1-\vec{s'}_o||\vec{s'}_2-\vec{s'}_o|} \equiv \hat{n'}_1 \cdot \hat{n'}_2.
\end{equation}
If the displacements are small, i.e., $\vec{q}_i \equiv (\vec{\Delta s}_i-\vec{\Delta s}_o)/|\vec{s}_i-\vec{s}_o|$ and $|\vec{q}_i| \ll 1$, then to first order in the small parameter
$|\vec{q}_i| $,
\begin{equation}
\hat{n'}_i \simeq \hat{n}_i +\vec{q}_i- (\vec{q}_i \cdot \hat{n}_i) \hat{n}_i,
\end{equation}
and consequently,
\begin{eqnarray}
&&F [\hat{n}_1,\hat{n}_2,\hat{n'}_1(\hat{n}_1,\vec{q}_1) ,\hat{n'}_2(\hat{n}_2,\vec{q}_2) ]
\equiv    \cos \gamma_{12} - \cos \gamma_{12}'
\nonumber \\
&& \hspace{35mm} \simeq   (\vec{q}_1 \cdot \hat{n}_1 + \vec{q}_2 \cdot \hat{n}_2) (\hat{n}_1 \cdot \hat{n}_2) - \vec{q}_1 \cdot \hat{n}_2 - \vec{q}_2 \cdot \hat{n}_1
\end{eqnarray}
gives the change in the angular separation between the two objects.

\subsection{Displacement maps}

Since a typical survey will observe several hundred thousand objects or more, we wish to condense the information in $F [\hat{n}_1,\hat{n}_2,\hat{n'}_1(\hat{n}_1,\vec{q}_1) ,\hat{n'}_2(\hat{n}_2,\vec{q}_2) ] $ for each pair of objects into some sort of average quantity.
To this end we parameterise
\begin{equation}
\vec{n}_i \doteq r_i(\cos \phi_i \sin \theta_i, \sin \phi_i \sin \theta_i, \cos \theta_i)^T,
\end{equation}
with $\phi_i \in [0,2 \pi)$ and $\theta_i \in [0,\pi)$, and similarly for $\hat{n'}_i$.
Then, for each set of $(\theta,\phi)$, we define
\begin{equation}
\label{eq:fff}
\langle F \rangle  (\theta,\phi) \equiv
\int_{r_{\rm min}}^{r_{\rm max}}  \rmd r  \ n_{\rm obj}(r) \int_{r_{\rm min}}^{r_{\rm max}}  \rmd r' \ n_{\rm obj}(r')
\int \frac{\rmd \Omega'}{4 \pi}   F,
\end{equation}
where $\rmd \Omega' =\sin \theta' \rmd \theta' \rmd \phi'$, and $n_{\rm obj}(r)$ is the number density distribution of observed objects (assumed here to depend only on $r$), normalised such that $\int_{r_{\rm min}}^{r_{\rm max}}  \rmd r  \ n_{\rm obj}(r) =1$.  Physically, this means that for each set of $(\theta,\phi)$, we first find the average $F$ value for each object found at these coordinates relative to all other objects in the sky, and then sum along the line-of-sight all objects at $(\theta,\phi)$.  The procedure results in a 2-dimensional map on the surface of a sphere, which can be further subjected to a decomposition
\begin{equation}
 \langle F \rangle  (\theta,\phi) = \sum_{\ell=0}^{\infty}\sum_{m=-\ell}^{\ell} a_{\ell m}Y_{\ell m}  (\theta,\phi),
\end{equation}
where $a_{\ell m} = \int d \Omega \  \langle F \rangle  (\theta,\phi) Y_{\ell m}^\star  (\theta,\phi)$,
in terms of the spherical harmonic functions $Y_{\ell m}( \theta, \phi)$.

It is simple to generalise this ``map-making'' procedure also to higher moments of $F$.  For example,
the second moment is
\begin{equation}
\label{eq:fffsquared}
\langle F^2  \rangle  (\theta,\phi) \equiv
\int_{r_{\rm min}}^{r_{\rm max}}  \rmd r  \ n_{\rm obj}(r) \int_{r_{\rm min}}^{r_{\rm max}}  \rmd r' \ n_{\rm obj}(r')
\int \frac{\rmd \Omega'}{4 \pi}   F^2,
\end{equation}
which has the spherical harmonic decomposition
\begin{equation}
\label{eq:sphericalf2}
\langle F^2 \rangle  (\theta,\phi) = \sum_{\ell=0}^{\infty}\sum_{m=-\ell}^{\ell} b_{\ell m}Y_{\ell m}  (\theta,\phi),
\end{equation}
and $b_{\ell m} = \int d \Omega \  \langle F^2 \rangle  (\theta,\phi) Y_{\ell m}^\star  (\theta,\phi)$ gives the corresponding multipole moments.

\subsection{Multipole vectors}

We defer the evaluation of $\langle F^n \rangle  (\theta,\phi)$ and the associated spherical harmonic multipole moments for specific cosmological models
to section~\ref{sec:examples}.  Suffice it to say here, however, that it is the {\it alignment} of the multipoles of  $\langle F\rangle  (\theta,\phi)$ and
$\langle F^2 \rangle  (\theta,\phi)$  that allows us to distinguish between radial expansion and rotation models.

To quantify the multipole alignments in a basis-independent fashion, we note that the $\ell$th multipole $T_\ell (\theta,\phi)$ can be expressed as~\cite{Copi:2003kt,Weeks:2004cz,Zunckel:2010yq}
\begin{equation}
\label{eq:multipolevectors}
T_\ell (\theta,\phi) \equiv  \sum_{m=-\ell}^{\ell} a_{\ell m} Y_{\ell m} (\theta,\phi) =
|\vec{n}|^{-\ell} \left[ \lambda_\ell   \prod^\ell_{i=1} (\vec{v}^{(\ell,i)} \cdot \vec{n}) + |\vec{n}|^{2} Q^{(\ell-2)}(\vec{n}) \right],
\end{equation}
where $\vec{n} \doteq (x,y,z)^T$, $\lambda_{\ell}$ is an $\ell$-dependent constant, and $Q^{(\ell-2)}(\vec{n})$ a $(\ell-2)$th degree polynomial
of $(x,y,z)$.  Importantly,  $\vec{v}^{(\ell,i)}$ are a set of cartesian (headless) vectors called the multipole vectors, which quantify the orientation of the multipole $T_\ell (\theta,\phi)$
in lieu of the spherical harmonic moments.  As an example, the dipole $T_1(\theta,\phi)$ can be translated into one single cartesian vector $\vec v^{(1,1)}$, with components
\begin{align}
	v^{(1,1)}_x &= -\sqrt{2} \ \mRe \left( a_{1,1} \right), &v^{(1,1)}_y =& -\sqrt{2} \ \mIm \left( a_{1,1} \right), &v^{(1,1)}_z =  a_{1,0} .
\end{align}
Higher multipoles vectors generally cannot be related to the spherical harmonic moments in such a straightforward manner.  However, as a guide, the real part of the spherical harmonic
$Y_{\ell m}(\theta,\phi)$ is spanned by $m$ vectors separated by an angle $\pi/m$ in the $xy$-plane, and $(\ell-m)$ vectors pointing in the $z$-direction.%
\footnote{A poor man's approach to computing the $\ell$th multipole vectors consists of first writing out the spherical harmonics $Y_{\ell m} (\theta,\phi)$ in
Cartesian coordinates, and evaluating the sum  $\sum_{m=-\ell}^{\ell} a_{\ell m} Y_{\ell m} (\theta,\phi)$ in equation~(\ref{eq:multipolevectors}).  This sum is then massaged into the form $r^{-\ell} [P^{(\ell)}(x,y,z) + r^2 Q^{(\ell-2)} (x,y,z)]$, where $P^{(\ell)}$ and $Q^{(\ell-2)}$ are polynomials of $(x,y,z)$ of degree $\ell$ and $(\ell-2)$ respectively,
and $r^2 = x^2+y^2+z^2$.
  In the final step we factorise
$P^{(\ell)}(x,y,z)$ into $\ell$ linear polynomials.  The resulting $i$th linear polynomial ($i = 1,\cdots,\ell$) is then of the form $v_x^{(\ell,i)} x + v_y^{(\ell,i)} y+v_z^{(\ell,i)} z$,
where the coefficients correspond to the Cartesian components of the $i$th vector $\vec{v}^{(\ell,i)}$ up to a constant normalisation factor.}

Once the multipole vectors have been determined, the question of alignment between any two multipoles can be addressed in terms of inner products of their corresponding
multipole (unit) vectors.
In this work, we shall be concerned mainly with the multipole vectors of $\langle F\rangle  (\theta,\phi)$ and $\langle F^2 \rangle  (\theta,\phi)$, denoted
\begin{align}
\hat v^{(l,i)}_{\langle F\rangle  (\theta,\phi)},&&
\hat v^{(l,i)}_{\langle F^2\rangle  (\theta,\phi)},
\end{align}
respectively.  Where numerical evaluation of these vectors is required, it is performed using the module of~\cite{Copi:2003kt} (although we have also tested it against the module published in~\cite{Weeks:2004cz}).

\section{Radial expansion versus rotation\label{sec:examples}}

In this section, we explore the differences between radial expansion and rotation models in terms of the alignment of the $\langle F\rangle  (\theta,\phi)$ and $\langle F^2 \rangle  (\theta,\phi)$
multipoles.   Generic predictions are summarised in section~\ref{sec:summary},   particularly table~\ref{tab:alignmentpredictions}.

\subsection{Radial expansion\label{sec:ltb}}

Lema\^{i}tre--Tolman--Bondi void models fall into this category.  Here, radial expansion displaces an object according to the displacement vector
\begin{equation}
\vec{\Delta s}_i =  \epsilon(s_i) \Delta t  \vec{s}_i,
\end{equation}
where $\epsilon (s_i)$ denotes the radial expansion rate which we assume to depend only on the distance $s_i \equiv |\vec{s_i}|$
to the origin, and $\Delta t$ is the time elapsed.  It follows that
\begin{equation}
\frac{F}{\Delta t} =  \frac{ \epsilon_1 - \epsilon_o}{r_1}
\left[( \hat{n}_1 \cdot \hat{n}_2)  \hat{n}_1-  \hat{n}_2  \right]  \cdot \vec{s}_o
+\frac{ \epsilon_2 - \epsilon_o}{r_2}
\left[( \hat{n}_1 \cdot \hat{n}_2) \hat{n}_2 -  \hat{n}_1 \right] \cdot \vec{s}_o,
\end{equation}
where we have used the shorthand notation $\epsilon_i \equiv \epsilon(s_i)$ and $r_i \equiv |\vec{n}_i|$.
 Without loss of generality we may choose the observer to be sitting on the $z$-axis, so that
$\vec{s}_o \doteq s_o(0,0,1)^{T}$, with $s_o \equiv |\vec{s}_o|$.

In order to evaluate analytically the moments~(\ref{eq:fff}) and~(\ref{eq:fffsquared}), we must first rewrite the expansion rate $\epsilon(s_i)$ as a function of
the distances $r_i$ and the angles $(\phi_i, \theta_i)$ measured by the observer.   Without specifying an exact form for $\epsilon(s_i)$, this is in general a non-trivial problem
because of the nonlinear dependence of $s_i = |\vec{n}_i +\vec{s}_o|$ on the said parameters, and  some form of expansion is usually necessary.  Given the symmetry of our set-up, we find it
convenient to expand $\epsilon(s_i)$ in terms of the Legendre series,
\begin{eqnarray}
\label{eq:legendre}
&& \epsilon(s_i) = \sum_{n=0}^\infty \tilde{\epsilon}_n (r_i,s_o) P_n (\mu_i), \nonumber \\
&& \tilde{\epsilon}_n(r_i,s_o) = \frac{2 n + 1}{2} \int _{-1}^{1} d \mu_i \ \epsilon(s_i) P_n (\mu_i),
\end{eqnarray}
where $\mu_i \equiv \vec{n}_i \cdot \vec{s}_o/s_o$, and $P_n(\mu)$ is a Legendre polynomial of degree $n$.
The advantage of the Legendre decomposition over, e.g., a Taylor expansion in $s_o/r_i$ is that the former can be used to explore both the $s_o \ll r_i$ (observer at a small distance from the origin) and the  $s_o \gg r_i$ (observer at a large distance from the origin) limits,
wherein the $n \geq 1$ Legendre moments of $\epsilon(s_i)$ characterise the anisotropic expansion rate {\it seen by the observer} as a result of their being removed from the origin.  As we shall see below, in some cases, this formulation even admits exact solutions valid for all $r_i$ and $s_o$ values.

Adopting the  expansion~(\ref{eq:legendre}), we find for the first moment $\langle F \rangle(\theta,\phi)$ an exact  solution
\begin{equation}
\label{eq:fltb}
\langle F \rangle  (\theta,\phi)  = \frac{2}{15}s_o \Delta t  \left \langle \frac{5\epsilon(s_o) -5 \tilde{\epsilon}_0(r)+\tilde{\epsilon}_2(r)}{r} \right \rangle
 \cos\theta,
\end{equation}
where $\langle \dots \rangle$ denotes averaging as per equation~(\ref{eq:fff})---in this particular instance,
this means averaging along the line-of-sight, weighted by the number density of observed objects---and we have omitted writing out the dependence of $\tilde{\epsilon}_n$ on $s_o$.
Clearly, the dependence of $\langle F \rangle  (\theta,\phi)$ on $\cos \theta$ indicates that $a_{1,0}$ is the only nonvanishing spherical harmonic moment.
This can be expressed equivalently in terms of multipole vectors as
\begin{equation}
\label{eq:ltbfdipole}
\hat{v}^{(1,1)}_{\langle F \rangle  (\theta,\phi)} = \hat{z},
\end{equation}
corresponding to a dipole that lies on the $z$-axis and connects to the origin.


An exact solution is not available for the second moment $\langle F^2 \rangle (\theta,\phi)$.  However,  decomposing $\langle F^2 \rangle (\theta,\phi)$ as per equation~(\ref{eq:sphericalf2}), symmetry arguments tell us that the only nonvanishing $b_{\ell m}$ entries are those with $m=0$.
An equivalent statement is that all multipole vectors are aligned with the $z$-axis, i.e.,
\begin{equation}
\label{eq:ltbf2multipole}
\hat{v}^{(\ell,i)}_{\langle F^2 \rangle  (\theta,\phi)} = \hat{z}, \qquad i=1,\ldots,\ell
\end{equation}
for $\ell \geq 1$,  This in turn leads to a generic prediction for radial expansion models: exact alignment between the $\langle F \rangle (\theta,\phi)$ dipole and all $\langle F^2 \rangle (\theta,\phi)$ multipoles, i.e.,
\begin{equation}
\label{eq:ltbprediction}
\hat{v}^{(1,1)}_{\langle F \rangle  (\theta,\phi)} \cdot \hat{v}^{(\ell,i)}_{\langle F^2 \rangle  (\theta,\phi)} =1, \qquad  i=1,\ldots,\ell,
\end{equation}
where $\ell \geq 1$, can be expected, which makes sense physically, given the only source of anisotropy in the model is the
fact that the observer is removed from the origin in the $z$-direction.

 For the aficionado, approximate solutions for $b_{\ell m}$ can be written down in the small ($s_o \ll r_i$) and the large ($s_o \gg r_i$) distance limits
by noting that, for expansion rates described by $\epsilon(s_i) \sim s_i^{-n}$, where $0<n \lwig 3$, the Legendre coefficients $\tilde{\epsilon}_n(r_i)$ behave roughly as
\begin{equation}
\label{eq:hierarchy}
|\tilde{\epsilon}_{n+1}(r_i)|  \sim  \xi |\tilde{\epsilon}_n(r_i)|,
\end{equation}
where $\xi \equiv s_o/r_i$ if $s_o \ll r_i$, and  $\xi \equiv r_i/s_o$ if $s_o \gg r_i$.  Thus, expanding the integrand $F^2$ in ``powers'' of $\xi$, we obtain  the leading order solutions
\begin{eqnarray}
&& b_{1,0} = -\frac{8}{75} \sqrt{\frac{\pi}{3}} s_o^2 \Delta t^2 \left \langle  \frac{[\epsilon(s_o) - \tilde{\epsilon}_0(r')] [3 r' \tilde{\epsilon}_1(r) + 5 r  \tilde{\epsilon}_1(r')]}{r r'^2} \right \rangle,
\nonumber \\
&& b_{2,0} = \frac{8}{45} \sqrt{\frac{\pi}{5}}  s_o^2 \Delta t^2\left \langle \left[ \frac{\epsilon(s_o) - \tilde{\epsilon}_0(r)}{r} \right]^2 \right \rangle , \nonumber \\
&& b_{3,0} = \frac{8}{75} \sqrt{\frac{\pi}{7}} s_o^2 \Delta t^2 \left \langle  \frac{[\epsilon(s_o) - \tilde{\epsilon}_0(r')] [3 r' \tilde{\epsilon}_1(r) + 5 r  \tilde{\epsilon}_1(r')]}{r r'^2} \right \rangle. 
\label{eq:radialb}
\end{eqnarray}
Observe that  $\langle F^2 \rangle (\theta,\phi)$ is dominated by the quadrupole moment $b_{2,0}$, while the dipole moment $b_{1,0}$ and octopole moment $b_{3,0}$ are both suppressed by $\xi$ relative to $b_{2,0}$.  For $\ell \geq 4$, $b_{\ell, 0} \sim \xi^{\ell-2} b_{2,0}$.

\subsection{Rotation\label{sec:origin}}

Without loss of generality we may choose the rotation axis to lie along the $y$-axis.   The corresponding displacement vector is then
\begin{equation}
\vec{\Delta s}_i =  \omega (\vec{s_i}) \Delta t {\bm A} \vec{s}_i,
\end{equation}
where $\omega (\vec{s_i})$ is the angular velocity, and
\begin{equation}
 {\bm A}  = \left( \begin{array}{ccc}
			0 & 0 & 1\\
			0 & 0 & 0 \\
			-1 & 0 & 0 \\ \end{array} \right)
\end{equation}
denotes rotation in the $xz$-plane.
This leads to
\begin{eqnarray}
\frac{F}{\Delta t}&= &
\frac{ \omega_1 - \omega_o}{r_1}
\left[( \hat{n}_1 \cdot \hat{n}_2)  \hat{n}_1-  \hat{n}_2  \right]  \cdot {\bm A} \vec{s}_o
+\frac{ \omega_2 - \omega_o}{r_2}
\left[( \hat{n}_1 \cdot \hat{n}_2) \hat{n}_2 -  \hat{n}_1 \right] \cdot {\bm A} \vec{s}_o  \nonumber \\
&& -\omega_1 \hat{n}_2 \cdot {\bm A} \hat{n}_1 - \omega_2 \hat{n}_1 \cdot {\bm A} \hat{n}_2 ,
\end{eqnarray}
where we have employed the shorthand notation $\omega_i \equiv \omega(\vec{s}_i)$.

\subsubsection{Observer at the origin}

Suppose firstly  that the observer sits at the origin so that $\vec{s}_0 \doteq (0,0,0)^T$.  Then, for any angular velocity model satisfying
\begin{equation}
\label{eq:omega}
\omega(r,\theta,\phi) = \omega(r,\pi-\theta,\phi) =
\omega(r,\theta,-\phi) = \omega(r,\theta,\pi-\phi),
\end{equation}
we find an exactly vanishing $\langle F \rangle  (\theta,\phi)$ and a nonzero $\langle F^2 \rangle  (\theta,\phi)$ given by
\begin{equation}
\label{eq:rot}
\langle F^2 \rangle  (\theta,\phi) =  A(\theta, \phi) \cos^2\theta + B(\theta,\phi) \cos^2 \phi \sin^2 \theta,
\end{equation}
with
\begin{eqnarray}
\label{eq:ab}
A(\theta, \phi)&\equiv& \Delta t^2 \langle [\omega(\vec{s})-\omega(\vec{s'})]^2 \sin^2\theta' \cos^2 \phi' \rangle
= A_1 + A_2(\theta, \phi), \nonumber \\
B(\theta, \phi)&\equiv& \Delta t^2\langle [\omega(\vec{s})-\omega(\vec{s'})]^2  \cos^2 \theta' \rangle
=B_1 + B_2(\theta, \phi),
\end{eqnarray}
where the residual dependence of $A$ and $B$ on $\theta$ and $\phi$ comes from  the angular dependence of the angular velocity $\omega(\vec{s})$.

For a model in which $\omega(\vec{s})$ depends only on the distance from the origin $s \equiv |\vec{s}|$, it is easy to show that $A_1=B_1$, while $A_2(\theta, \phi)=B_2(\theta,\phi)=0$ .  In this case, the only nonzero spherical harmonic moments of $\langle F^2 \rangle (\theta,\phi)$ are
\begin{align}
\label{eq:rotdecomposition}
b_{2,0} = \frac{2}{3} \sqrt{\frac{\pi}{5}} (2 A_1 - B_1) ,
&& b_{2,\pm 2}= \sqrt{\frac{2 \pi}{15}} B_1 ,
\end{align}
corresponding to a quadrupole aligned with the axis of rotation, i.e.,
\begin{equation}
\label{eq:rotf2quadrupole}
\hat{v}^{(2,i)}_{\langle F^2 \rangle  (\theta,\phi)} = \hat{y}, \qquad i=1,2.
\end{equation}
In fact, all odd multipole moments of $\langle F^2 \rangle(\theta,\phi)$ are exactly vanishing as long as $\omega(\vec{s})$ satisfies
equation~(\ref{eq:omega}).
If  $A_2(\theta,\phi)$ and $B_2(\theta,\phi)$ should be nonzero because of an angular dependence in $\omega(\vec{s})$, e.g.,
the rotating region exhibits a cylindrical symmetry so that $\omega(\vec{s}) = \omega(s_x^2 + s_z^2)$, then we expect the spherical harmonic
 decomposition to generate additional terms for even values of $\ell$ and~$m$.

\subsubsection{Off-axis and off-centre observers\label{sec:off}}

Consider now a more realistic rotation model, in which the observer sits not at the origin, but at a location given by $\vec{s}_o \doteq (-s_{ox},s_{oy},0)^T$.
Here, the combination of a vanishing $s_{ox}$ and a nonzero $s_{oy}$ denotes an observer who is away from the origin, but still
on the rotation axis.  The orthogonal combination of $s_{ox} \neq 0$ and $s_{oy}=0$ indicates an off-axis observer sitting on the equatorial $xz$-plane intersecting the origin.

To simplify the problem, we assume the angular velocity $\omega(\vec{s}_i)$ to depend only on the distance from the origin, i.e., $\omega(\vec{s}_i) = \omega(s_i)$.  Then, a Legendre
decomposition of $\omega(s_i)$,
\begin{equation}
\label{eq:legendre2}
 \omega(s_i) = \sum_{i=0}^\infty \tilde{\omega}_n (r_i,s_{ox},s_{oy}) P_n (\mu_i),
\end{equation}
yields an exact solution for the first moment $\langle F \rangle  (\theta,\phi)$:
\begin{equation}
\label{eq:frot}
\langle F \rangle  (\theta,\phi)  = \frac{1}{15}s_{ox} \Delta t \left \langle \frac{10 [\omega(s_o) -\tilde{\omega}_0(r)]-5 (s_{ox}/s_o) \tilde{\omega}_1(r)
-\tilde{\omega}_2(r)}{r} \right \rangle
 \cos\theta,
\end{equation}
where $s_o^2 \equiv s_{ox}^2 +s_{oy}^2$, and we have omitted writing out the dependence of $\tilde{\omega}_n$ on $s_{ox}$ and $s_{oy}$.
This expression shows that being merely off-centre (i.e, $s_{oy} \neq 0$) does not generate a nonzero $\langle F \rangle  (\theta,\phi)$, but being off-axis (i.e., $s_{ox} \neq 0$) does, and
the angular signature is exactly like that for radial expansion (see equation~(\ref{eq:fltb})).  In other words, we have again a dipole aligned with the $z$-axis,
\begin{equation}
\hat{v}^{(1,1)}_{\langle F \rangle  (\theta,\phi)} = \hat{z},
\end{equation}
or, equivalently, a nonzero $a_{1,0}$, corresponding to the direction of the observer's motion on the $xz$-plane.

An exact analytical solution does not exist for  $\langle F^2 \rangle (\theta,\phi)$.  However, a brute-force evaluation of the integral~(\ref{eq:fffsquared}) does enable us to establish that
$b_{1,0}=0$, meaning that  the dipole of  $\langle F^2 \rangle (\theta,\phi)$ lies in general on the $xy$-plane.  It is also interesting to examine the nonvanishing elements of $b_{1m}$ in the limits of small and large observer-to-origin distances.  Using the empirical fact that
$|\tilde{\omega}_{n+1}(r)| \sim \xi |\tilde{\omega}_n(r)|$, where $\xi \equiv s_o/r_i$ or $r_i/s_o$, with $s_o^2 \equiv s_{ox}^2 + s_{oy}^2$ (cf equation~(\ref{eq:hierarchy})), we find the leading order solutions:
\begin{eqnarray}
&&b_{1,\pm 1} =\pm \frac{2}{3} \sqrt{\frac{2 \pi}{3}} \Delta t^2\left \{
 s_{ox} \left \langle \frac{[\tilde{\omega}_0(r')-\tilde{\omega}_0(r)][\tilde{\omega}_0(r') - \omega(s_o)]}{r'} \right \rangle
 \right. \nonumber \\
&& \hspace{30mm} \left. +  \frac{2}{5} \left(
\frac{2 s_{ox} \pm  i  s_{oy}}{s_o} \right)
\left  \langle [\tilde{\omega}_0(r')-\tilde{\omega}_0(r)]\tilde{\omega}_1(r') \right \rangle\right \}
\end{eqnarray}
for $s_o \ll r_i$, and
\begin{eqnarray}
&&b_{1,\pm1} = \pm \frac{2}{3} \sqrt{\frac{2 \pi}{3}} \Delta t^2
 \left \{
s_{ox} \left \langle \frac{[\tilde{\omega}_0(r')-\tilde{\omega}_0(r)][\tilde{\omega}_0(r') - \omega(s_o)]}{r'} \right \rangle \right.
\nonumber \\
&&  \left.  + \frac{4}{25}s_{ox}^2 \left( \frac{s_{ox} \pm  i s_{oy}}{s_o} \right)
  \left \langle \frac{[2 r' \tilde{\omega}_1(r) - 5 r \tilde{\omega}_1(r')]  [\omega(s_o) - \tilde{\omega}_0(r')]}{r r'^2} \right \rangle
  \right \}
\end{eqnarray}
for $s_o \gg r_i$.
From these expressions, we see that although the dipole lies on the $xy$-plane, its precise orientation is strongly dependent on the angular velocity model as well as on the precise location of the observer.  Furthermore, unlike the dipole of $\langle F \rangle (\theta, \phi)$ which can be understood to indicate the direction of the observer's motion around the rotation axis, the orientation of  $\langle F^2 \rangle (\theta, \phi)$'s dipole has no clear physical interpretation; for instance, it does not  in general point back to the origin.  Importantly, however, the dipoles of $\langle F \rangle (\theta, \phi)$ and $\langle F^2 \rangle (\theta, \phi)$ are {\it always} orthogonal to each other in rotation models such that
\begin{equation}
\hat{v}^{(1,1)}_{\langle F \rangle  (\theta,\phi)} \cdot \hat{v}^{(1,1)}_{\langle F^2 \rangle  (\theta,\phi)}  = 0.
\end{equation}
 This is in contrast with the case of radial expansion, where the two dipoles are always exactly aligned (see equation~(\ref{eq:ltbprediction})).

\begin{figure}[t]
\begin{center}
\includegraphics{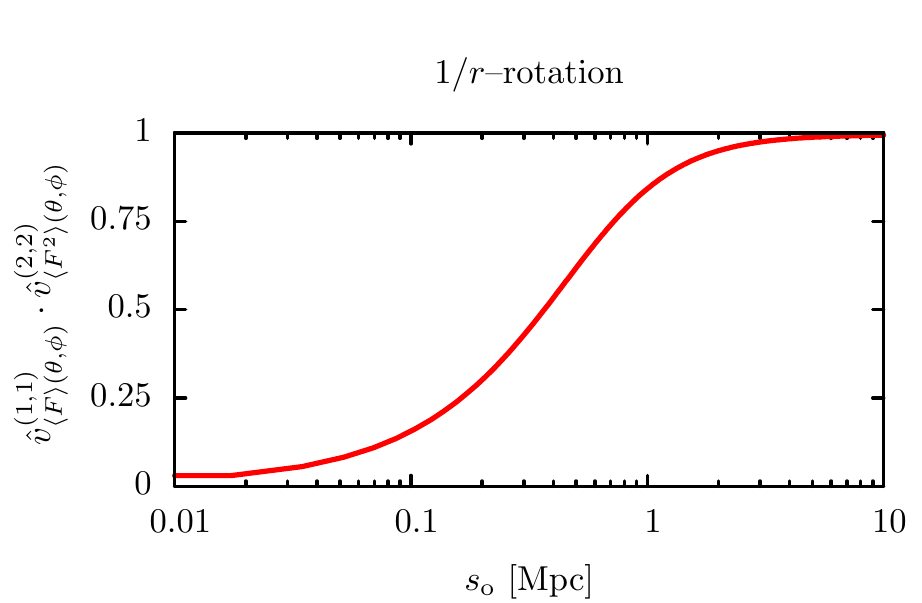}
\end{center}
\caption{The inner product between the first vector of the  ${\left<F\right>(\theta,\phi)}$ dipole and the second vector of ${\left<F^2\right>(\theta,\phi)}$ quadrupole, as a function of the  observer's distance from the origin in the $1/r$-rotation model.    In this example, rotation axis is chosen to lie along the $y$-axis, while the observer lives
on the coordinates $x=y=z$.
The orientation of the ${\left<F^2\right>(\theta,\phi)}$ quadrupole makes a rapid transition from being orthogonal to being parallel to the  ${\left<F\right>(\theta,\phi)}$ dipole as the observer moves away from the origin.   }\label{fig:inpvsr}
\end{figure}

The quadrupole of $\langle F^2 \rangle (\theta, \phi)$ likewise exhibits different limiting behaviours.  In the small distance $s_{ox},s_{oy} \ll r_i$ limit, the leading order $b_{2,m}$ moments  are simply those given in equation~(\ref{eq:rotdecomposition}) (for an observer sitting at the origin), with
\begin{equation}
\label{eq:ab2}
A_1 = B_1= \frac{1}{3} \Delta t^2 \langle[ \tilde{\omega}_0(r) - \tilde{\omega}_0(r')]^2 \rangle ,
\end{equation}
indicating that both quadrupole vectors are aligned with the rotation axis as per equation~(\ref{eq:rotf2quadrupole}).
This result remains true even as the observer moves further away from the origin ($s_{oy} \gg r_i$), as long as he/she stays close to the rotation axis, i.e., $s_{ox} \ll s_{oy}$.  However, if the observer is significantly removed from the rotation axis ($s_{ox} \sim s_{oy}$ or $s_{ox} \gg s_{oy}$), then the large distance $s_{ox},s_{oy} \gg r_i$ limit sees $b_{2,0}$ dominating over $b_{2,\pm 2}$, with
\begin{equation}
\label{eq:far}
b_{2,0} = \frac{8}{45} \sqrt{\frac{\pi}{5}} s_{ox}^2 \Delta t^2\left \langle
\frac{[\omega (s_o) - \tilde{\omega}_0(r)]^2}{r^2}
\right \rangle ,
\end{equation}
resulting in quadrupole vectors
\begin{equation}
\label{eq:offf2quadrupole}
\hat{v}^{(2,i)}_{\langle F^2 \rangle  (\theta,\phi)} = \hat{z}, \qquad i=1,2
\end{equation}
that are now aligned with the $z$-axis.  This ``flipping'' of the quadrupole alignment is clearly sourced by the increasing importance of the observer's own motion in the $xz$-plane:
when close to the origin, the dominant effect seen by the observer is the difference between the rotation rates of the various objects along the rotation axis (and hence the dependence on
$\tilde{\omega}_0(r) - \tilde{\omega}_0(r')$ in equation~(\ref{eq:ab})).
When far away from the origin, however, the observer perceives this difference less than he/she would perceive the difference between their own motion in the $z$-direction and the motion of these objects
(and hence the dependence on $\omega(s_o) - \tilde{\omega}_0(r')$ in equation~(\ref{eq:far})).  Where the flipping occurs is model-dependent.
We therefore argue that the quadrupole of  $\langle F^2 \rangle (\theta, \phi)$ is perhaps not very useful as a diagnostic of rotation models.
Figure~\ref{fig:inpvsr} illustrates this flipping by way of the inner product of $\hat{v}^{(1,1)}_{\langle F \rangle  (\theta,\phi)} \cdot \hat{v}^{(2,2)}_{\langle F^2 \rangle  (\theta,\phi)}$
for the $1/r$-rotation model (in which $\omega_(s_i) \propto 1/s_i$; see section~\ref{sec:anismodels} for details).

The octopole of $\langle F^2 \rangle (\theta, \phi)$ has in general $b_{3,0}=b_{3,\pm 2}=0$,  and displays the same ``flipping'' feature as the quadrupole.  On the one hand,
for small displacements ($s_{ox},s_{oy} \ll r_i$) or an on-axis observer ($s_{ox } \ll s_{oy}$), the non-vanishing multipoles are, to leading order,
\begin{eqnarray}
&& b_{3,\pm1} = \pm \frac{2}{15} \sqrt{\frac{\pi}{21}}  \left(\frac{s_{ox} \pm  i 3 s_{oy}}{s_o} \right) \Delta t^2 \left \langle \tilde{\omega}_1(r') [\tilde{\omega}_0(r') - \tilde{\omega}_0(r)]  \right \rangle, \nonumber \\
&& b_{3,\pm 3} = \pm \frac{2}{3} \sqrt{\frac{\pi}{35}}\left(  \frac{s_{ox} \pm i  s_{oy}}{s_o} \right) \Delta t^2 \left \langle \tilde{\omega}_1(r') [\tilde{\omega}_0(r') - \tilde{\omega}_0(r)]  \right \rangle ,
\end{eqnarray}
indicating an octopole spanned by the unit vectors
\begin{eqnarray}
\label{eq:octopole1}
&& \hat{v}^{(3,i)}_{\langle F^2 \rangle  (\theta,\phi)} = \hat{y}, \qquad i=1,2, \nonumber \\
&&  \hat{v}^{(3,3)}_{\langle F^2 \rangle  (\theta,\phi)} = \frac{1}{s_{o}} (-s_{ox} ,s_{oy},0)^T,
\end{eqnarray}
on the $xy$-plane.  On the other hand, for a distant off-axis observer ($s_{oy} \gg r_i$),
$b_{3,\pm 1}$ dominates the solution, with
\begin{eqnarray}
&& b_{3,\pm 1} = \pm \frac{8}{75} \sqrt{\frac{\pi}{21}} s_{ox}^2  \left(  \frac{s_{ox} \pm  i s_{oy}}{s_o} \right)  \Delta t^2 \nonumber \\
&& \hspace{30mm}  \times \left \langle  \frac{[ \omega(s_o) -\tilde{\omega}_0(r')]   [3 r'  \tilde{\omega}_1 (r) +5 r \tilde{\omega}_1(r')] }{r r'^2} \right \rangle,
\end{eqnarray}f
and multipole vectors
\begin{eqnarray}
\label{eq:octopole2}
&& \hat{v}^{(3,i)}_{\langle F^2 \rangle  (\theta,\phi)} = \hat{z}, \qquad i=1,2, \nonumber \\
&&  \hat{v}^{(3,3)}_{\langle F^2 \rangle  (\theta,\phi)} = \frac{1}{s_{o}} (-s_{ox} ,s_{oy},0)^T.
\end{eqnarray}
Comparing equations~(\ref{eq:octopole1}) and~(\ref{eq:octopole2}), we see that even though two of the three multipole vectors flip from the $y$-direction to the $z$-direction depending on the observer's distance to the origin, one appears to always lie on the $xy$-plane and consistently point back to the origin.   Thus, we have a model-independent signature: at least one vector of the octopole of $\langle F^2 \rangle  (\theta,\phi)$  satisfies the orthogonality condition
\begin{equation}
\hat{v}^{(1,1)}_{\langle F \rangle  (\theta,\phi)} \cdot \hat{v}^{(3,3)}_{\langle F^2 \rangle  (\theta,\phi)} =0.
\end{equation}
This  is in contrast with the predictions for radial expansion models~(\ref{eq:ltbprediction}), where all $\langle F^2 \rangle  (\theta,\phi)$ multipole vectors are exactly aligned with the
$\langle F \rangle  (\theta,\phi)$ dipole.

As in the case of radial expansion, higher moments ($\ell \geq 4$) of $\langle F^2 \rangle  (\theta,\phi)$ are suppressed according to $b_{\ell, m} \sim \xi^{\ell-2} b_{2, m}$.

\subsection{Summary\label{sec:summary}}

Table~\ref{tab:alignmentpredictions} summarises the generic, model-independent predictions for the inner products of the $\langle F \rangle  (\theta,\phi)$ dipole vector with
the dipole, quadrupole and octopole vectors of  $\langle F^2 \rangle  (\theta,\phi)$, in  radially expanding and rotating anisotropic universe.
Importantly, by observing the alignment or orthogonality of the dipoles of $\langle F \rangle  (\theta,\phi)$ and $\langle F^2 \rangle  (\theta,\phi)$, we should in principle be able to distinguish
decisively between these two classes of anisotropic models.  Further distinctions  can be gleaned from the octopole vectors of $\langle F^2 \rangle  (\theta,\phi)$, which again
show opposite behaviours for radially expanding and rotating universes.

{\renewcommand{\arraystretch}{2}
\begin{table}[t]
\begin{center}
\begin{tabular}{ll|cc}
 & Observable   & LTB void &Rotation  \\
 \hline
A & $ \hat v^{(1,1)}_{\left<F  \right>(\theta,\phi)}\cdot \hat v^{(1,1)}_{\left<F^2\right>(\theta,\phi)} $  & $ {  1  }  $  & $ {  0 }  $ \\
B & $ \hat v^{(1,1)}_{\left<F  \right>(\theta,\phi)}\cdot \hat v^{(2,\{1,2\})}_{\left<F^2\right>(\theta,\phi)} $  & $ {  1 }  $  & $ {  0 \to 1,0 \to 1 } $   \\
C & $ \hat v^{(1,1)}_{\left<F  \right>(\theta,\phi)}\cdot \hat v^{(3,\{1,2,3\})}_{\left<F^2\right>(\theta,\phi)} $  & $ {  1  }  $  & $ {0 \to 1,0 \to 1,0 }  $ \\
D & $ \hat v^{(1,1)}_{\left<F^2\right>(\theta,\phi)}\cdot \hat v^{(2,\{1,2\})}_{\left<F^2\right>(\theta,\phi)} $  & $ { 1 }  $  & $ {  {\rm md} \to 0  }  $   \\
E &  $ \hat v^{(1,1)}_{\left<F^2\right>(\theta,\phi)}\cdot \hat v^{(3,\{1,2,3\})}_{\left<F^2\right>(\theta,\phi)} $  & $ {  1 }  $  & $ {{\rm md} \to 0, {\rm md} \to 0 ,{\rm md} }  $
\end{tabular}
\end{center}
\caption{Generic predictions for various inner products in LTB and rotation models.  In the latter case the observer is assumed to be both off-axis and off-centre.   The notation $x \to y$ denotes the inner product can change from $x$ to $y$ when going from small to large displacements from the centre, while ``md'' indicates a model-dependent number between 0~and~1.}
\label{tab:alignmentpredictions}
\end{table}
}
We assess in section~\ref{sec:forecast}  the actual detectability of a selection of radial expansion and rotation models,  assuming the survey specifications of GAIA and beyond.

\section{Numerical analysis}\label{sec:V}

In order to assess the potential of RTC to resolve the multipole alignments reported in table~\ref{tab:alignmentpredictions} and thereby distinguish between different classes of anisotropic models, we perform numerical simulations of the observables for the models and survey under question.
We take the specifications of the GAIA quasar survey~\cite{Jordi:2005te} as a guideline, and adopt in particular its expected object count of 500.000 QSOs on the full sky.  For the survey angular resolution, we consider a range of numbers starting from GAIA's nominal resolution of $\sigma_{\rm res} \sim 10~\mu$as (micro-arcsecond).  The time gap between successive observations is also let free to vary.

\subsection{Simulation scheme}
Our numerical computation%
\footnote{Our code is available for download at \url{http://web.physik.rwth-aachen.de/download/valkenburg/Rotation/}.}
consists of the following steps:
\begin{enumerate}
\item Distribute 500.000 quasars on uncorrelated random positions in the sky, each with a redshift drawn from a probability distribution  based on the redshift distribution of  77.430 quasars in the SDSS database~\cite{Abazajian:2008wr}.

\item Displace the quasars according to the anisotropic model under question and the desired time gap between observations.   Likewise for the observer.

\item Induce gaussian errors on the angular positions by adding a random angular displacement to each quasar drawn
 from a normal distribution with a zero mean and a standard deviation corresponding to the angular resolution of the survey.

\item Induce a random displacement for each quasar according to a gaussian peculiar velocity distribution with zero mean and standard deviation $110$~km s$^{-1}$~(e.g.,~\cite{Kashlinsky:2009dw}) multiplied by the time between observations.%
\footnote{Reference~\cite{Kashlinsky:2009dw} gives the root-mean-square velocity as $v_{\rm rms} \sim 250 \ (100 \ h^{-1} \ {\rm Mpc}/d) \ {\rm km \ s^{-1}}$ for 
$d> 50  \to 100 \ h^{-1} \ {\rm Mpc}$ in the concordance $\Lambda$CDM model, where $d$ is the distance from the observer, and our choice of $v_{\rm ms} = 110\ {\rm km \ s}^{-1}$ corresponds to $d \sim 230 \ h^{-1} \ {\rm Mpc}$.  We have opted not to model the distance-dependence of the peculiar velocity distribution in this work because all observables considered here involve an integration along the line-of-sight, which would negate much of the distance-modelling.}

\item Bin the displacement information on pixels using Healpix~\cite{Gorski:2004by}, and bin in redshift space.
\item Perform the integrals~\eqref{eq:fff}~and~\eqref{eq:fffsquared} by weighted sums over the bins.
\item Compute the spherical harmonic moments of  $\langle F \rangle  (\theta,\phi)$ and $\langle F^2 \rangle  (\theta,\phi)$ using Healpix.
\item Calculate the corresponding multipole vectors using the module of~\cite{Copi:2003kt}.
\item Finally, evaluate inner products of the multipole vectors.
\end{enumerate}

Each iteration yields a set of inner products representing the observable outcome of a particular realisation of an inherently random measurement.  We therefore repeat the scheme
$N$~times, each time with a different set of random seeds, in order to obtain for each inner product of interest a probability distribution that automatically incorporates the measurement and the sampling errors intrinsic to the survey configuration.

One iteration of the above scheme takes of order seconds on a standard 2~GHz processor,
so that we can easily produce $N=1000$ random realisations for any one chosen anisotropic model.  The resulting $N$~sets of inner products are saved in a format compatible with {\sc GetDist}, which we then use to generate plots of the probability distributions. {\sc GetDist} is a tool for analysing Monte-Carlo chains, and is supplied in all standard releases
of  {\sc CosmoMC}~\cite{Lewis:2002ah}.

\subsection{Criteria for detection}

Our first goal is to determine if any inner product can be detected at all, i.e., to see if there is a peak in its probability distribution.  Since we use normalised vectors, each
inner product $x$ obeys strictly  $0 \leq x \leq 1$.  In the absence of a  signal (because, for instance, of an undefined multipole in the cosmological model under question), any measurement of $x$ must be the result of noise only, and will be distributed equally between 0 and 1.  Its expectation value will be $\left<x\right>=0.5$, with variance
\begin{align}
\label{eq:sigma}
\sigma^2 = & \frac{\int_0^1 (x-0.5)^2 dx }{ \int_0^1 dx} = 1/12,
\end{align}
or, equivalently, $\sigma \simeq 0.289$.  Hence, we define a detected peak as a probability distribution with a standard deviation $\sigma$ significantly smaller than $0.289$.

As  the standard deviation~$\sigma$ is estimated from a finite sample, it comes with its own sampling error $\sigma_{\sigma}$ given by~\cite{Nakamura:2010zzi}
\begin{align}
\sigma_{\sigma}=\frac{\sigma}{\sqrt{2N}}, \label{eq:varianceofvariance}
\end{align}
where $N$ is the number of samples. From this it follows that for $N=1000$, the relative error on $\sigma$ is  2.2\%. Henceforth, we shall always take 1000 realisations to estimate the errors on the inner products. If we assume all processes to be random and the resulting probability distributions purely gaussian, then undefined inner products will have a standard deviation $0.282<\sigma<0.295$ 68\% of the time, and a $5\sigma$-detection of a peak in a distribution is made if  $\sigma<0.257$.

\section{Forecast of detectability}\label{sec:forecast}

\subsection{Compared universes\label{sec:anismodels}}

We consider the following list of radial expansion and rotation models, as well as the isotropic $\Lambda$CDM model which forms our  null hypothesis.  In each model
the observer is put on the coordinates $s_{ox}=s_{oy}=s_{oz}$, but at various distances from the origin.

\paragraph{$\Lambda$CDM}
For the null hypothesis, we take the isotropic standard cosmological model with a flat spatial geometry and the parameters $\Omega_{\Lambda} = 0.7$, $\Omega_{\rm matter} = 0.3$, and $H_0 = 73$~km s$^{-1}$ Mpc$^{-1}$. In this model the quasars have no angular displacement besides that due to random peculiar velocities and the gaussian uncertainty in the angular position measurement.
They do however have a redshift drift.

\paragraph{Large local void}
For the radial expansion scenario, we consider a LTB void model with a density profile referred to as ``Void A'' in~\cite{Biswas:2010xm} and adopt the best-fit parameters from the same reference.  This model consists of a low-density region described by a two-parameter density profile embedded in a spatially curved dust-filled universe. The spherically symmetric
low-density region extends to redshift $z=1.07$ in radius, and has a central underdensity of $(\rho_{\rm in} - \rho_{\rm out})/\rho_{\rm out} = -0.668$.  The global spatial curvature is positive with $\Omega_k = 0.2$, and the  age of the universe in this model is 17.5~Gyr.

\paragraph{$1/r^2$-rotation}
The $1/r^2$-rotation model corresponds to objects in stable circular orbits around a point mass at the origin.
All objects within 2~Gpc of the origin of rotation, including the observer if the observer is sufficiently close to the axis of rotation, revolve about the rotation axis with angular velocity
$\omega(s) = 2.3 \times 10^{-10} \left(10~{\rm Mpc}/ s\right)^{2}$~radians per year, where $s$ is the distance of the object from the origin.
We cut off to solid body rotation (i.e., constant $\omega$) at $s<10$~Mpc, and  to no rotation ($\omega=0$) at $s>2$~Gpc. This means that on the equatorial plane at 10~Mpc from the origin, objects have physical velocities of $7\times10^{-3}c$, or $2\times 10^{3}$~km s$^{-1}$. The rotating region is embedded in the previously described $\Lambda$CDM universe.

\paragraph{$1/r$-rotation}
The $1/r$-rotation model represents another configuration of objects in stable circular orbits in a dark matter halo. All objects within 2~Gpc of the origin, including the observer,
move angular velocity with $\omega(s) = 6.7 \times 10^{-11}\left(10~{\rm Mpc}/ s\right)$~radians per year around the rotation axis.
Again, we cut off to  solid body rotation (constant $\omega$)  at $s<10$~Mpc, and to no rotation ($\omega=0$) at $s>2$~Gpc. This means that all objects on the equatorial plane between 10~Mpc  and 2~Gpc from the origin move with a constant tangential velocity of $2\times10^{-3} c$, or $6\times 10^{2}$ km s$^{-1}$. Again, the rotating region is embedded in the $\Lambda$CDM universe described above.

\paragraph{Solid body rotation}
The solid body rotation model assumes a solid sphere with a radius of 1~Gpc rotating with an angular velocity of $\omega =10^{-12}$~radians per year. Objects on the equatorial plane at 1~Gpc from the origin therefore have physical velocities $\sim 3\times10^{-3}c$, or $10^{3}$~km s$^{-1}$.  This rotating sphere is embedded in a standard $\Lambda$CDM universe.

\subsection{Results}

\subsubsection{Extreme anisotropy}\label{subsec:extreme}

We consider first a hypothetical extreme situation in which the observer lives
\begin{itemize}
\item at 1~Gpc from the origin for the local void and the $1/r$-rotation model,
\item at 35~Mpc for $1/r^2$-rotation, and
\item at 690~Mpc for solid rotation,
\end{itemize}
and is able to make two observations each with a resolution of 1~$\mu$as (a factor of 10 better than GAIA's nominal resolution) and a gap of 1000 years between them.
Besides the improbable waiting time between observations, these configurations are unrealistic in that the observer inevitably sees a dipole in the cosmic microwave background (CMB) induced by the observer's own proper motion that is much larger than what is currently observed.  Nonetheless, because such extreme configurations afford us strong signals, they serve to verify our analytical understanding and to identify the most easily accessible observables.

\begin{figure}[t]
\begin{center}
\includegraphics[width=0.95\textwidth]{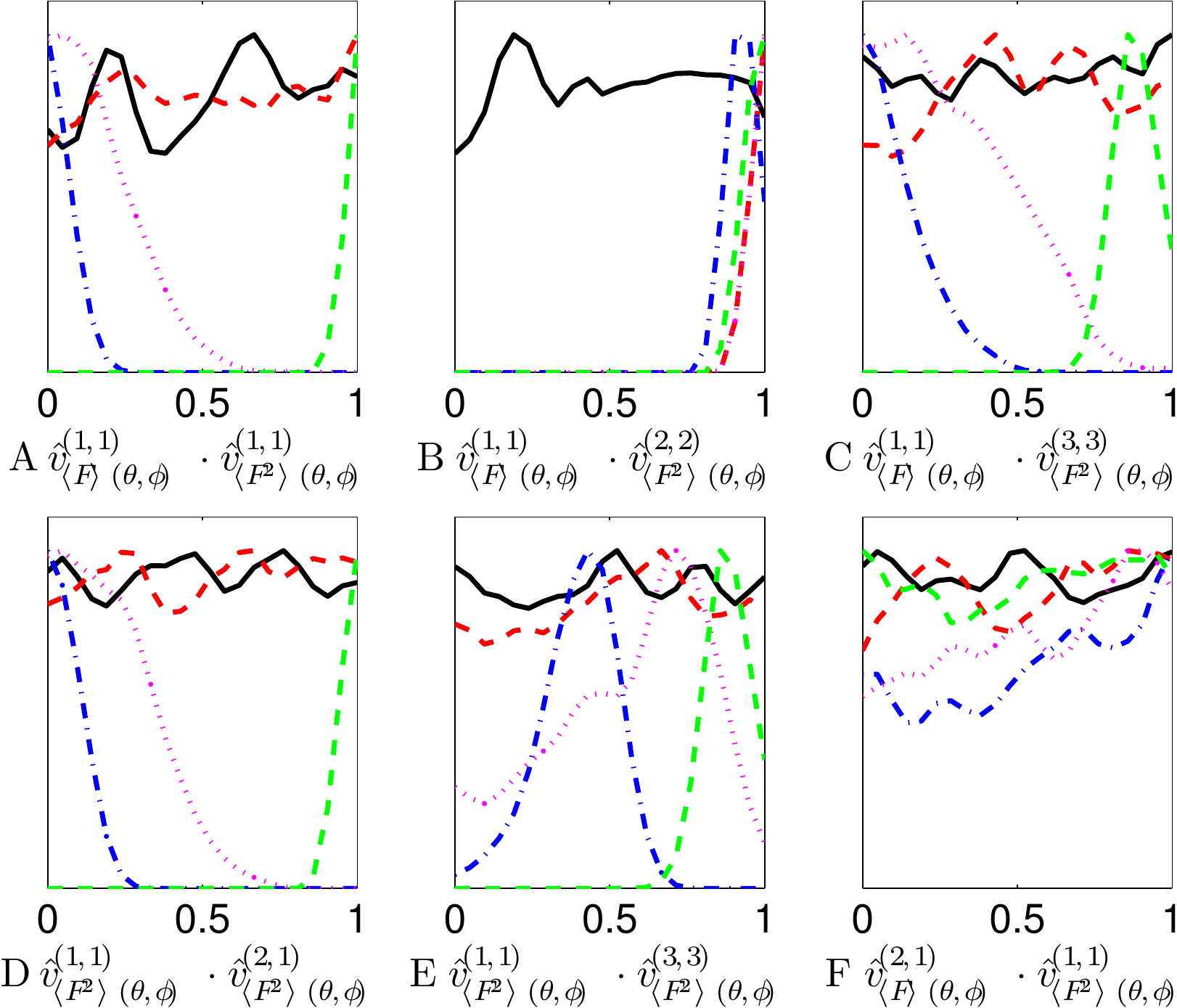}
\end{center}
\caption{Probability distributions, based on 1000 mock universes, for a selection of inner products assuming an observer who waits 1000 years between two observations each with a resolution of 1~$\mu$as. Solid black lines denote the $\Lambda$CDM model, dashed red the $1/r^2$-rotation model (observer at 35~Mpc from the origin), dashed-dotted blue
$1/r$-rotation (observer at 1~Gpc), dotted magenta solid body rotation (observer at 690~Mpc), and dashed green the large local void model (observer at 1~Gpc).
In all models except $\Lambda$CDM and $1/r^2$-rotation, the observer sees a sky that is in disagreement with current observations (see text for discussion).  However, these configurations serve as an example with clear signals and confirm the analytical results of table~\ref{tab:alignmentpredictions}.
}
\label{fig:inprods2013}
\end{figure}

Figure~\ref{fig:inprods2013} shows the probability distributions for a selection of inner products listed in
table~\ref{tab:alignmentpredictions} for
the five specific models discussed in section~\ref{sec:anismodels}, and, for reference, one inner product that has no defined signal in any of the models,
namely, observable~F, $ \hat v^{(2,1)}_{\left<F  \right>(\theta,\phi)}\cdot \hat v^{(1,1)}_{\left<F^2\right>(\theta,\phi)} $, the inner product of the nonexistent $\left<F \right>$ quadrupole and the $\left< F^2 \right>$ dipole.  As expected, our null hypothesis, the $\Lambda$CDM model, produces no discernible peaks in the probability distributions for any of the inner products.
Their corresponding standard deviations, shown in
table~\ref{tab:inprods2013},  likewise do not pass our ``detection threshold'' of $\sigma < 0.257$.  The same is true also for the reference observable~F, which is undefined in all five considered models and consequently generates no signal.

\begin{table}[t]
\begin{stabular}{ll|ccccc}
&   Observable & $\Lambda$CDM & Large local void & Solid rotation & $1/r$--rotation & $1/r^2$--rotation\\
&                      &                            & $s_o = 1$ Gpc   & $s_o = 690$ Mpc & $s_o = 1$ Gpc & $s_o = 35$ Mpc \\
 \hline
A & $ \hat v^{(1,1)}_{\left<F  \right>(\theta,\phi)}\cdot \hat v^{(1,1)}_{\left<F^2\right>(\theta,\phi)} $  & $  0.504 \pm 0.292  $  & $ {\bf  0.995 \pm 0.003 }  $  & $ {\bf  0.173 \pm 0.126 }  $  &$ {\bf  0.046 \pm 0.037 }  $ &$  0.511 \pm 0.289  $ \\
B & $ \hat v^{(1,1)}_{\left<F  \right>(\theta,\phi)}\cdot \hat v^{(2,2)}_{\left<F^2\right>(\theta,\phi)} $  & $  0.509 \pm 0.290  $  & $ {\bf  0.960 \pm 0.002 }  $  & $ {\bf  0.981 \pm 0.005 }  $ & $ {\bf  0.919 \pm 0.005 }  $   &$ {\bf  0.981 \pm 0.005 }  $ \\
C & $ \hat v^{(1,1)}_{\left<F  \right>(\theta,\phi)}\cdot \hat v^{(3,3)}_{\left<F^2\right>(\theta,\phi)} $  & $  0.502 \pm 0.289  $  & $ {\bf  0.873 \pm 0.049 }  $  & $ {\bf  0.298 \pm 0.200 }  $  & $ {\bf  0.124 \pm 0.102 }  $  & $  0.496 \pm 0.278  $ \\
D & $ \hat v^{(1,1)}_{\left<F^2\right>(\theta,\phi)}\cdot \hat v^{(2,1)}_{\left<F^2\right>(\theta,\phi)} $  & $  0.493 \pm 0.283  $  & $ {\bf  0.971 \pm 0.011 }  $  & $ {\bf  0.212 \pm 0.144 }  $  & $ {\bf  0.069 \pm 0.048 }  $  & $  0.508 \pm 0.290  $\\
E & $ \hat v^{(1,1)}_{\left<F^2\right>(\theta,\phi)}\cdot \hat v^{(3,3)}_{\left<F^2\right>(\theta,\phi)} $  & $  0.503 \pm 0.288  $  & $ {\bf  0.870 \pm 0.053 }  $  & $ {\bf  0.560 \pm 0.242 }  $ & $ {\bf  0.396 \pm 0.117 }  $  & $  0.514 \pm 0.288  $ \\
F & $ \hat v^{(2,1)}_{\left<F  \right>(\theta,\phi)}\cdot \hat v^{(1,1)}_{\left<F^2\right>(\theta,\phi)}  $  & $  0.510 \pm 0.288  $   &$  0.509 \pm 0.293  $            & $  0.542 \pm 0.289  $            & $  0.547 \pm 0.295  $            & $  0.512 \pm 0.284  $ \\
\hline
\end{stabular}
\caption{The central values and standard deviations corresponding to the probability distributions presented in figure~\ref{fig:inprods2013}.
Where a standard deviation is smaller than 0.257 (i.e., a value for $\sigma$ that is less than $5\sigma_\sigma$ away from the standard deviation expected from a purely random signal, where the sampling error $\sigma_\sigma = \sigma/\sqrt{2 N}$ is non-zero because we produce only $N=1000$ samples), we typeset the result in boldface to indicate that an inner product is
detectable in the given configuration.}
\label{tab:inprods2013}
\end{table}

Contrastingly, an observer living in a large local void
in this extreme configuration would clearly measure every well-defined inner product.  In particular, using only the information in the dipoles of $\left<F\right>$ and $\left<F^2\right>$ (observable A), measurement of a sharp peak  at $ \hat v^{(1,1)}_{\left<F  \right>(\theta,\phi)}\cdot \hat v^{(1,1)}_{\left<F^2\right>(\theta,\phi)} =1$ would immediately allow this observer to rule out anisotropic models due to local rotation.
A similarly strong signal is in principle available also for the inner product of the $\left<F\right>$ dipole and the $\left<F^2\right>$ quadrupole (observable B).
However, because both radial expansion and rotation models produce identical signatures in the large displacement limit for this observable (see table~\ref{tab:alignmentpredictions}),
a detected peak at~$\sim 1$ cannot be used as a basis to discriminate between the two.  Inner products involving the octopole of~$\left< F^2 \right>$ (observables C and E) could likewise be measured to a good accuracy.

Of the three rotation models considered here, only the $1/r$-rotation and the solid rotation models lead to detectable signals in all applicable cases.   In both models observable~B, the inner product of the $\langle F \rangle$ dipole and the $\langle F^2 \rangle$ quadrupole, appears to be especially well-determined, although, as discussed above, this signal does not serve to distinguish between radial expansion and rotation models.  Of the remaining observables, those involving the octopole of $\left<F^2\right>$ (observables~C and E) are significantly less detectable than those formed from its dipole or quadrupole (observables A and D).

Lastly, the $1/r^2$-rotation model manifests itself only through observable B, and produces a signal that is identical to that of the local void.

In summary, the most easily accessible observable appears to be B, the inner product of the $\left<F\right>$ dipole and the $\left<F^2\right>$ quadrupole, followed by observable~A,  the inner product of the $\left<F\right>$ and the $\left<F^2\right>$ dipoles.  The latter is necessary for the distinction between radial expansion and rotation models.

\subsubsection{Observationally consistent anisotropy}\label{sec:consistent}

We now place the observer at a new location at which their proper velocity approximates $\sim 500$ km s$^{-1}$, a number roughly consistent with the velocity of the Local Group inferred from the measured CMB dipole~\cite{Hinshaw:2008kr} albeit at the high end.   This amounts to putting the observer
\begin{itemize}
\item at $690$~Mpc for solid body rotation, and
\item at $35$~Mpc for all other models,
\end{itemize}
again with the assumption of $s_{ox}=s_{oy}=s_{oz}$.
We focus on the detectability of observables~A and~B as functions of the
survey's angular resolution and the time gap between successive observations.

\begin{figure}
\begin{center}
{\em Detectability of $ \hat v^{(1,1)}_{\left<F \right>(\theta,\phi)}\cdot \hat v^{(2,2)}_{\left<F^2\right>(\theta,\phi)} $ }\\
\resizebox{\textwidth}{!}{
\begin{tabular}{l}
\includegraphics[width=6cm,trim=0 1cm 1.7cm 0,clip=true]{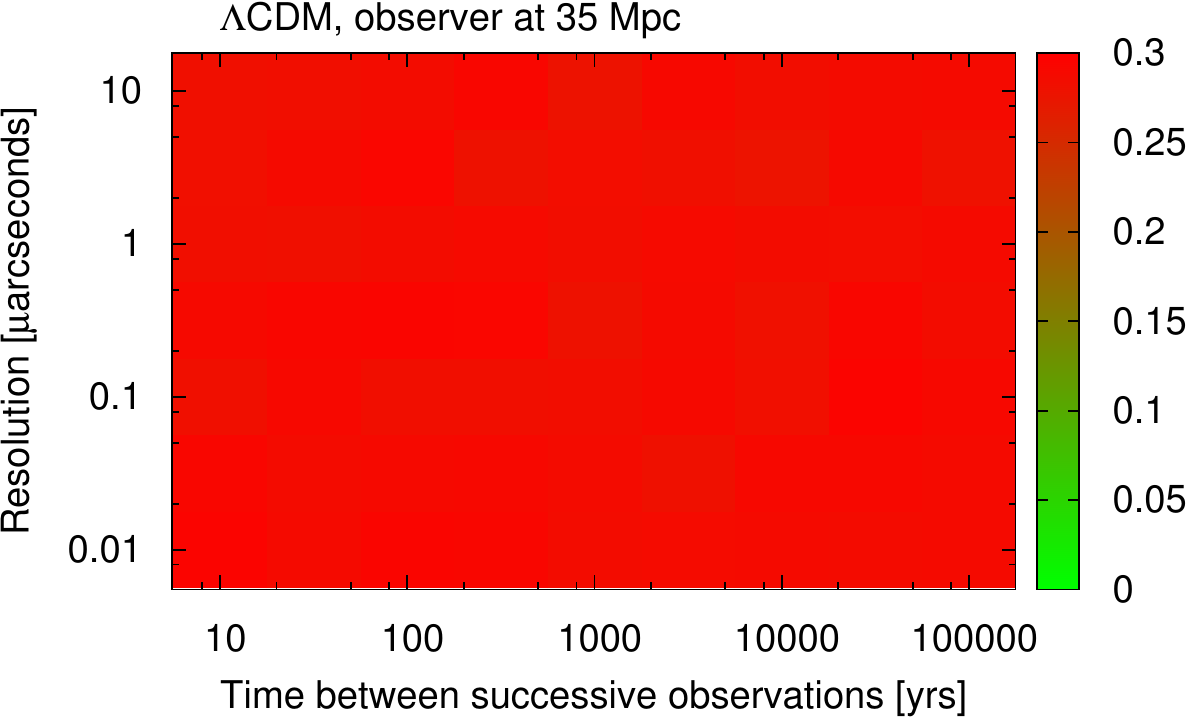}
\hspace{-0.25cm}
\includegraphics[width=6.2cm,trim=1.4cm 1cm 0cm 0,clip=true]{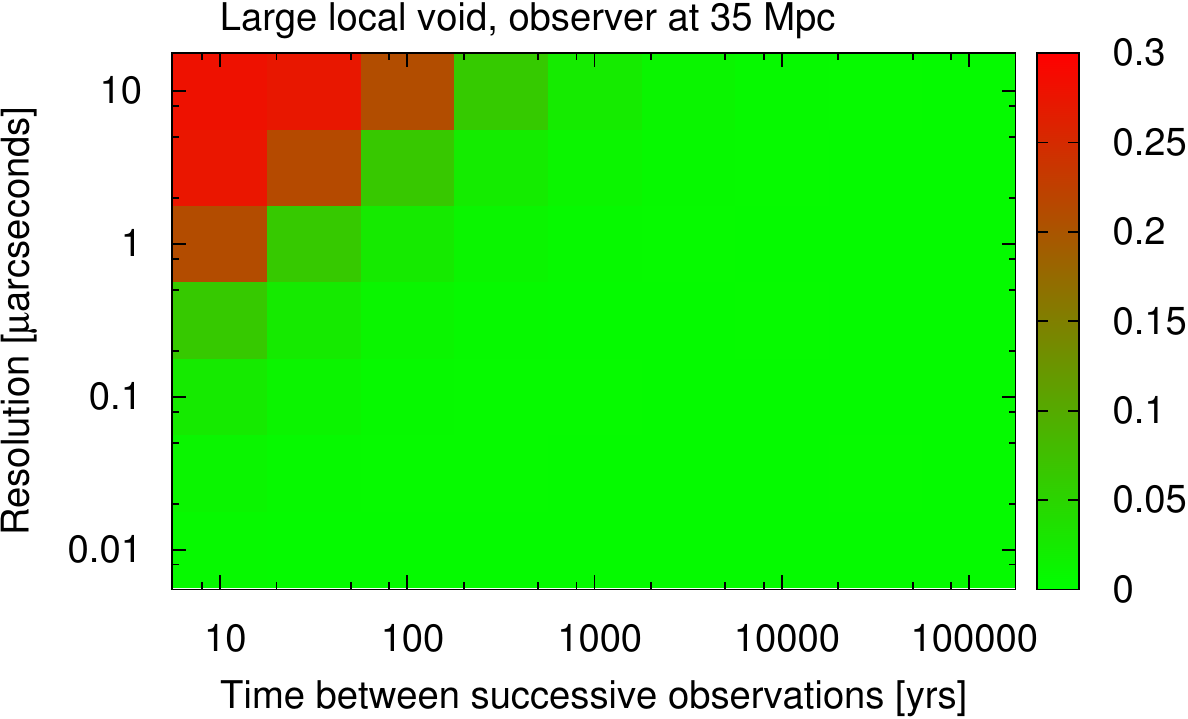}\vspace{-0.35cm}\\
\hspace{-0.1cm}\imagetop{
\includegraphics[width=6cm,trim=0 1cm 1.7cm 0,clip=true]{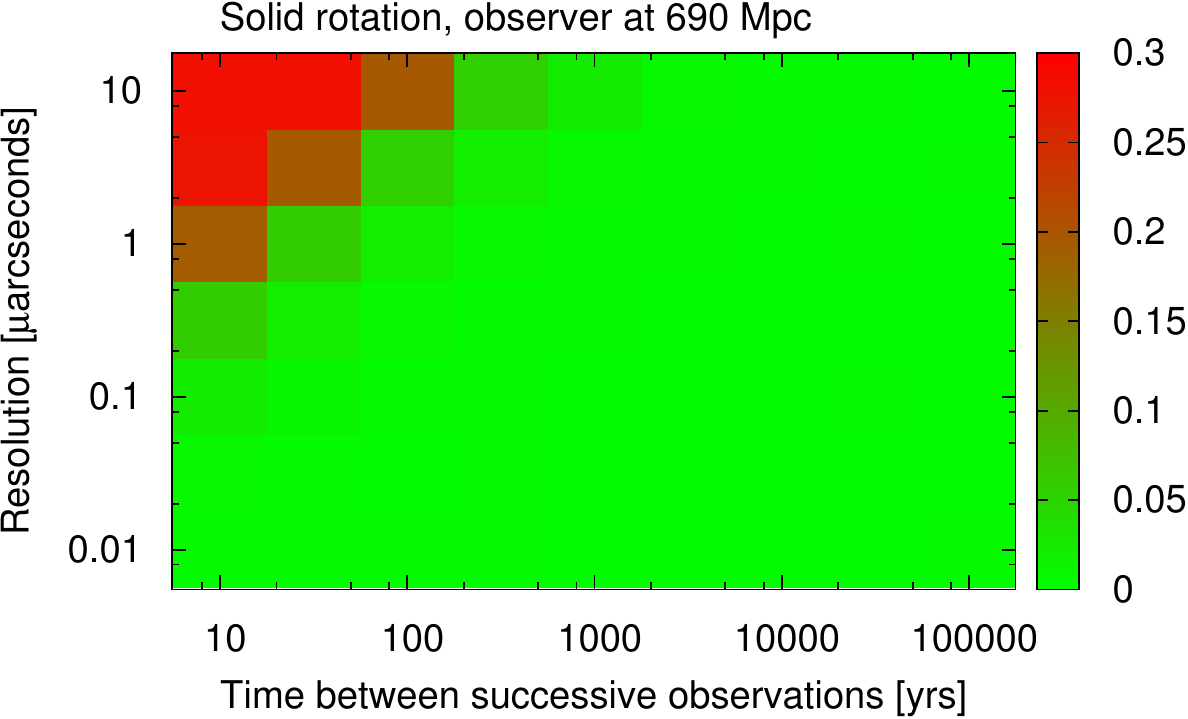}}
\hspace{-0.25cm}\imagetop{
\includegraphics[width=6.2cm,trim=1.4cm 0cm 0cm 0,clip=true]{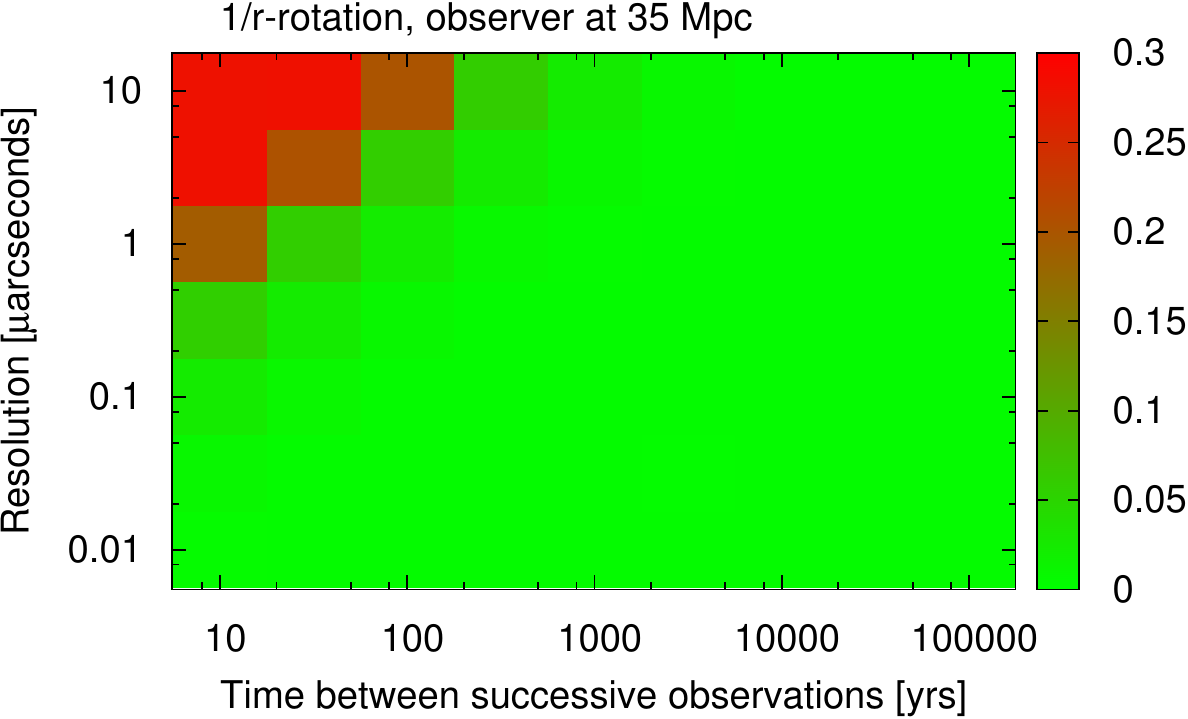}}\vspace{-0.35cm}\\
\includegraphics[width=6cm,trim=0 0cm 1.7cm 0,clip=true]{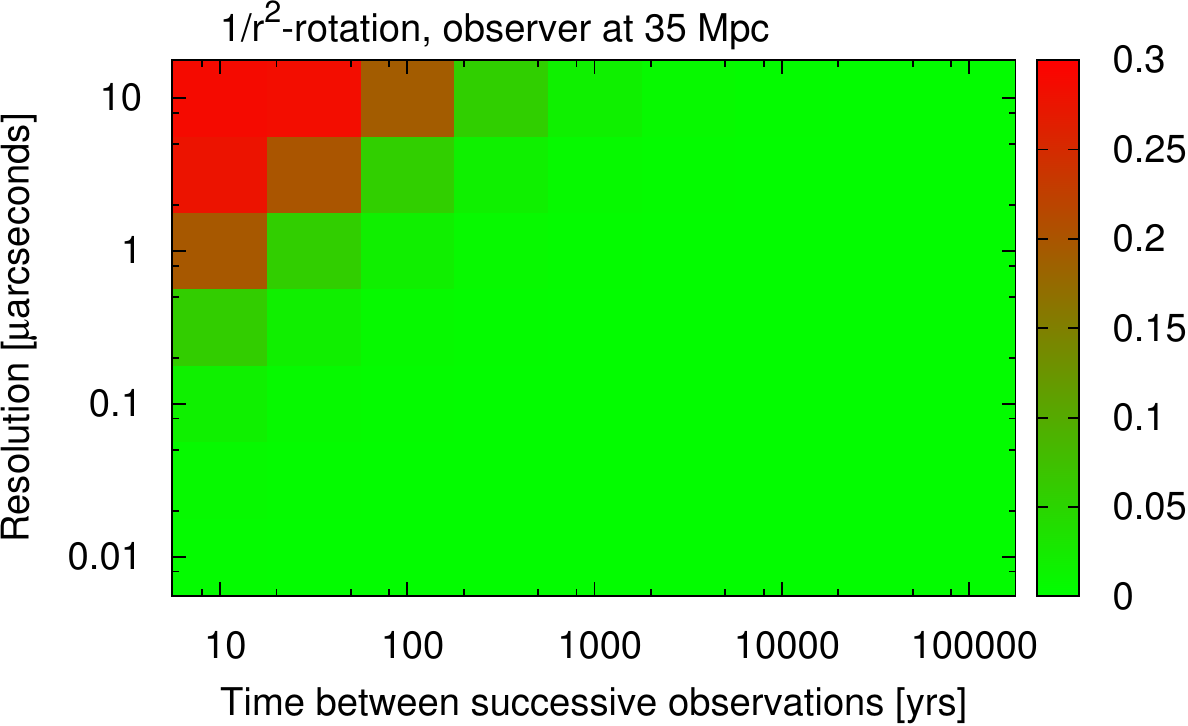}
\end{tabular} }
\end{center}
\caption{Standard deviation in the inner product B, $ \hat v^{(1,1)}_{\left<F \right>(\theta,\phi)}\cdot \hat v^{(2,2)}_{\left<F^2\right>(\theta,\phi)} $, and hence the detectability of that inner product, for various combinations of survey resolution (vertical axis) and time between successive observations (horizontal axis).  The observer always sits  at 35~Mpc from the origin, except in the case of solid body rotation in which $s_o=690$~Mpc.
For this inner product, all anisotropic models under consideration predict $ \hat v^{(1,1)}_{\left<F \right>(\theta,\phi)}\cdot \hat v^{(2,2)}_{\left<F^2\right>(\theta,\phi)}  = 1$, and are hence indistinguishable from one another.  Nonetheless, detection of this observable becomes possible after a 10-year time gap at a survey resolution of 1~$\mu$as.
For an inner product that distinguishes between different classes of anisotropic models, see figure~\ref{fig:plane_detectable}.}
\label{fig:2dplanebigsignal}
\end{figure}

Figure~\ref{fig:2dplanebigsignal} shows the detectability of observable B, $ \hat v^{(1,1)}_{\left<F \right>(\theta,\phi)}\cdot \hat v^{(2,2)}_{\left<F^2\right>(\theta,\phi)} $, for the five different models.   Clearly,  in all models, a higher survey resolution has the same effect as a longer gap between observations, while in rotation models, a longer time-gap is in addition one-to-one degenerate with a larger rotation velocity.  Detection of observable~B becomes possible in all anisotropic models after 10~years at a survey angular resolution of 1~$\mu$as.
 However, as already discussed in section~\ref{subsec:extreme}, the inner product  $\hat v^{(1,1)}_{\left<F \right>(\theta,\phi)}\cdot \hat v^{(2,2)}_{\left<F^2\right>(\theta,\phi)}$ is an inconclusive
 observable, because the expected value is  1 in both a large local void and rotation models in the large observer displacement limit.

\begin{figure}
\begin{center}
{\em Detectability of $ \hat v^{(1,1)}_{\left<F\right>(\theta,\phi)}\cdot \hat v^{(1,1)}_{\left<F^2\right>(\theta,\phi)} $ }\\
\resizebox{\textwidth}{!}{
\begin{tabular}{l}
\includegraphics[width=6cm,trim=0 1cm 1.7cm 0,clip=true]{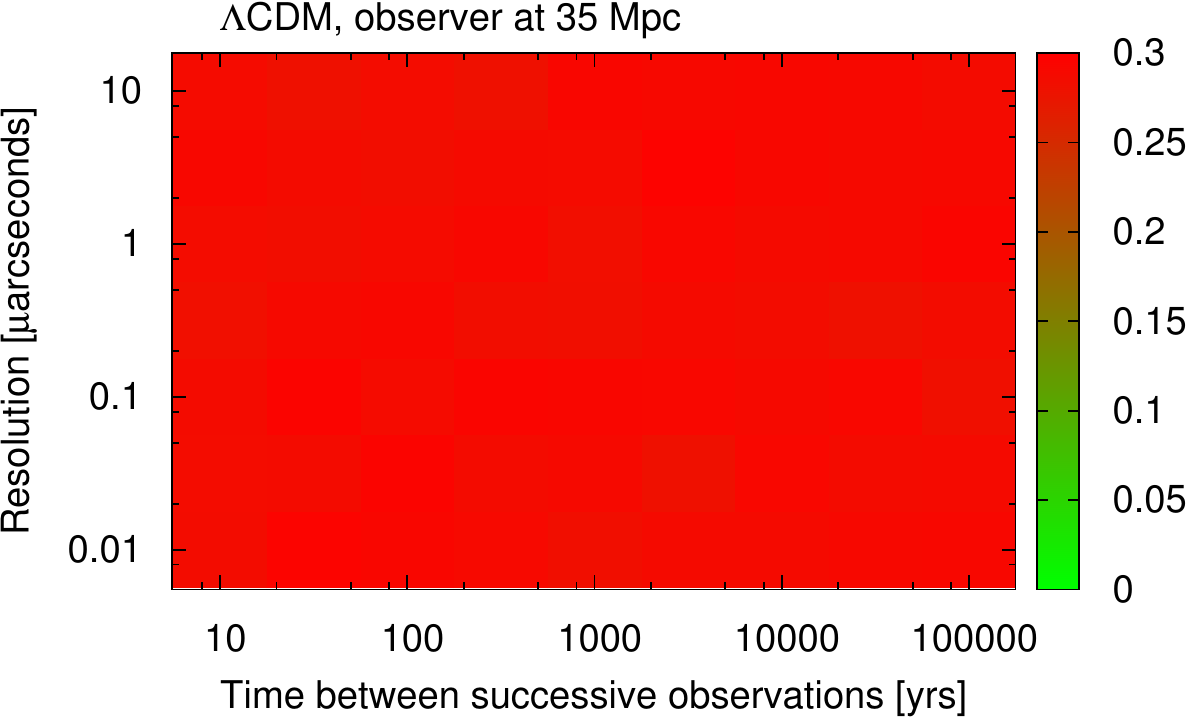}
\hspace{-0.25cm}
\includegraphics[width=6.2cm,trim=1.4cm 1cm 0cm 0,clip=true]{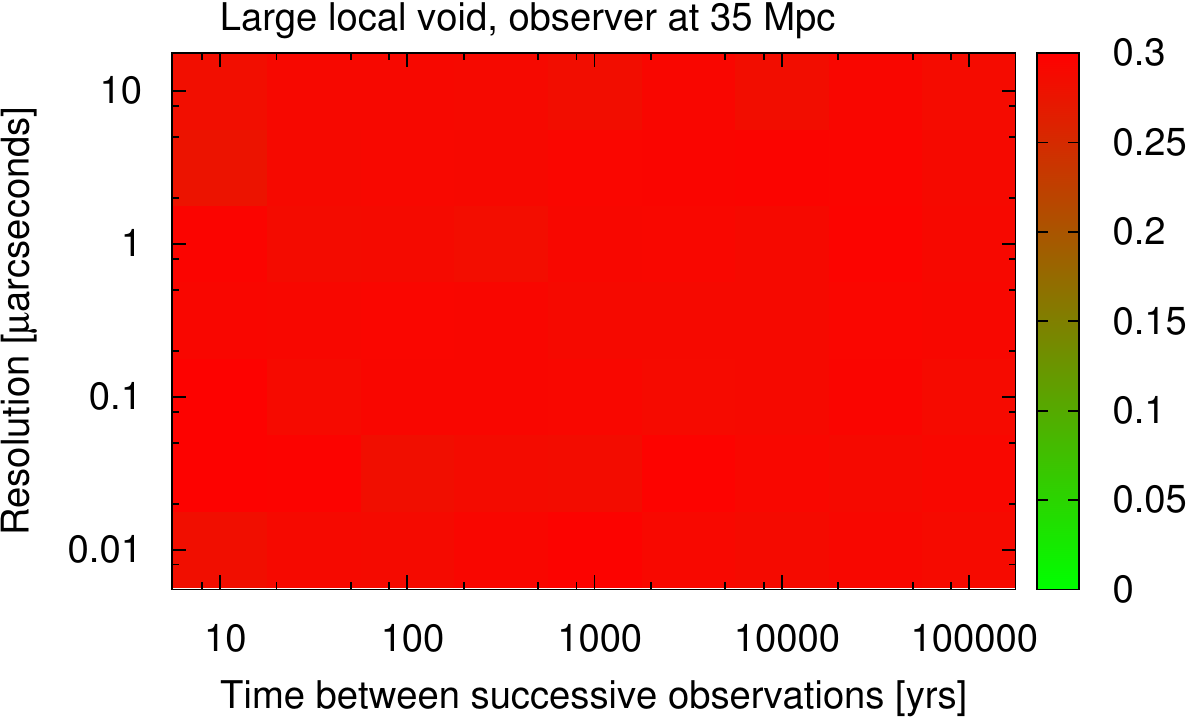}\vspace{-0.35cm}\\
\hspace{-0.1cm}\imagetop{
\includegraphics[width=6cm,trim=0 1cm 1.7cm 0,clip=true]{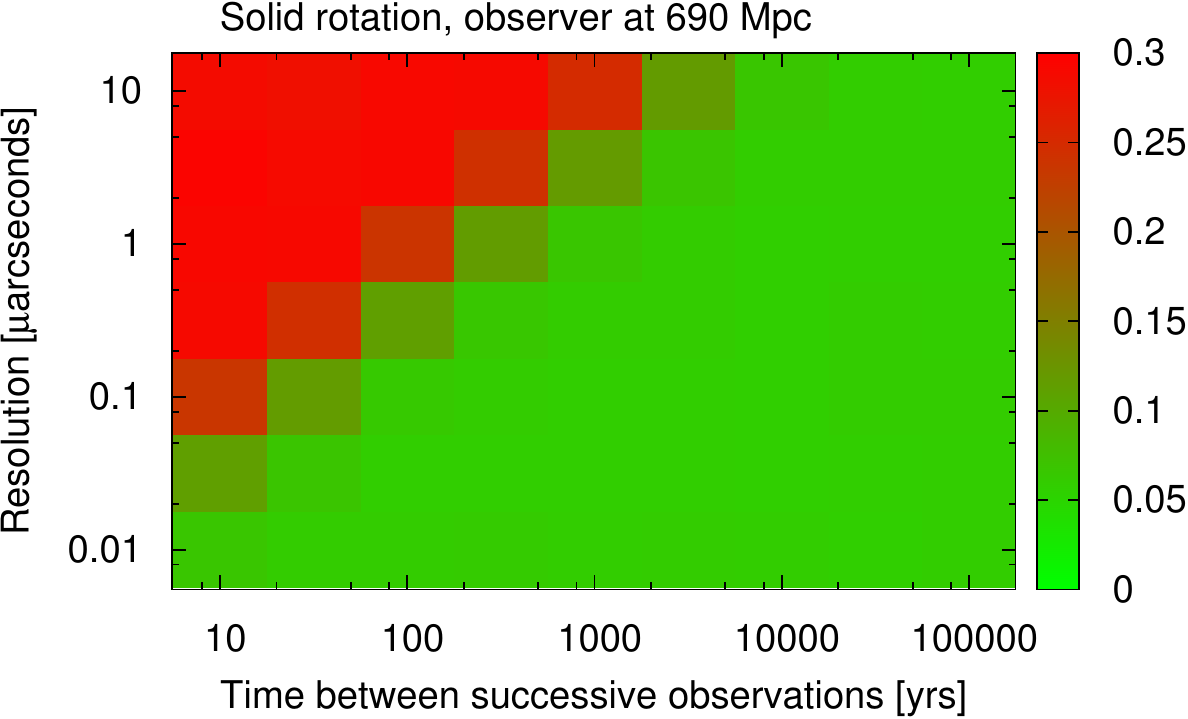}}\hspace{-0.25cm}
\imagetop{
\includegraphics[width=6.2cm,trim=1.4cm 0cm 0cm 0,clip=true]{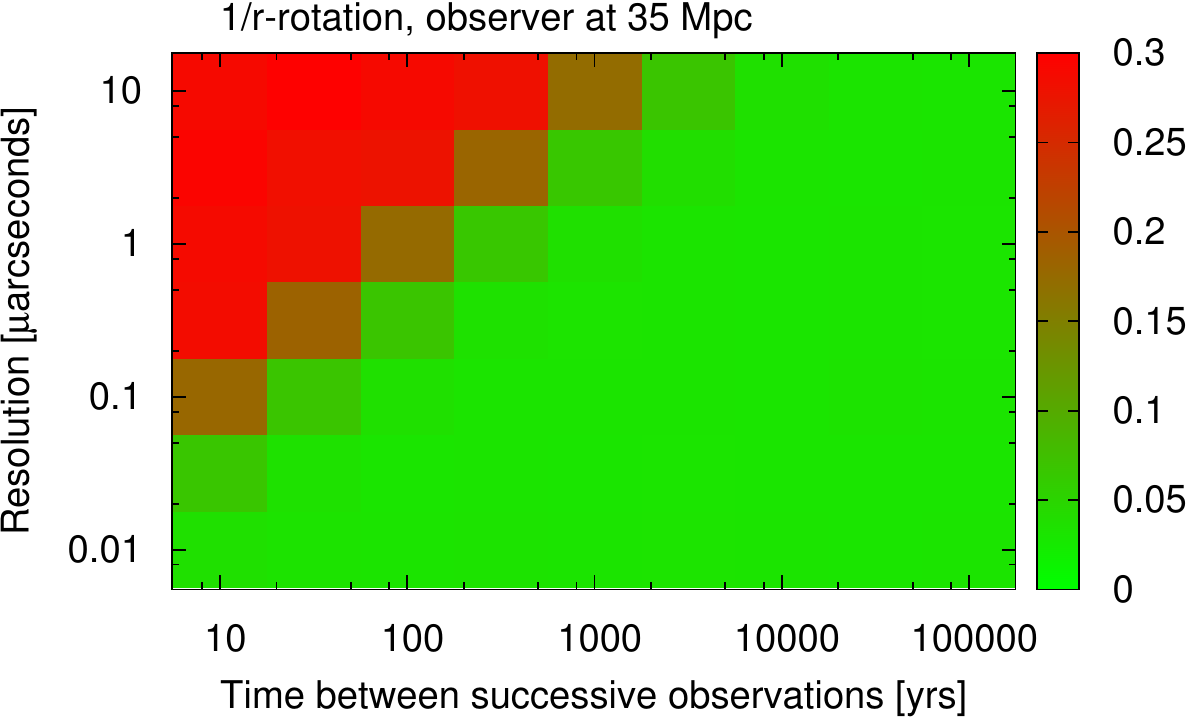}}\vspace{-0.35cm}\\
\includegraphics[width=6cm,trim=0 0cm 1.7cm 0,clip=true]{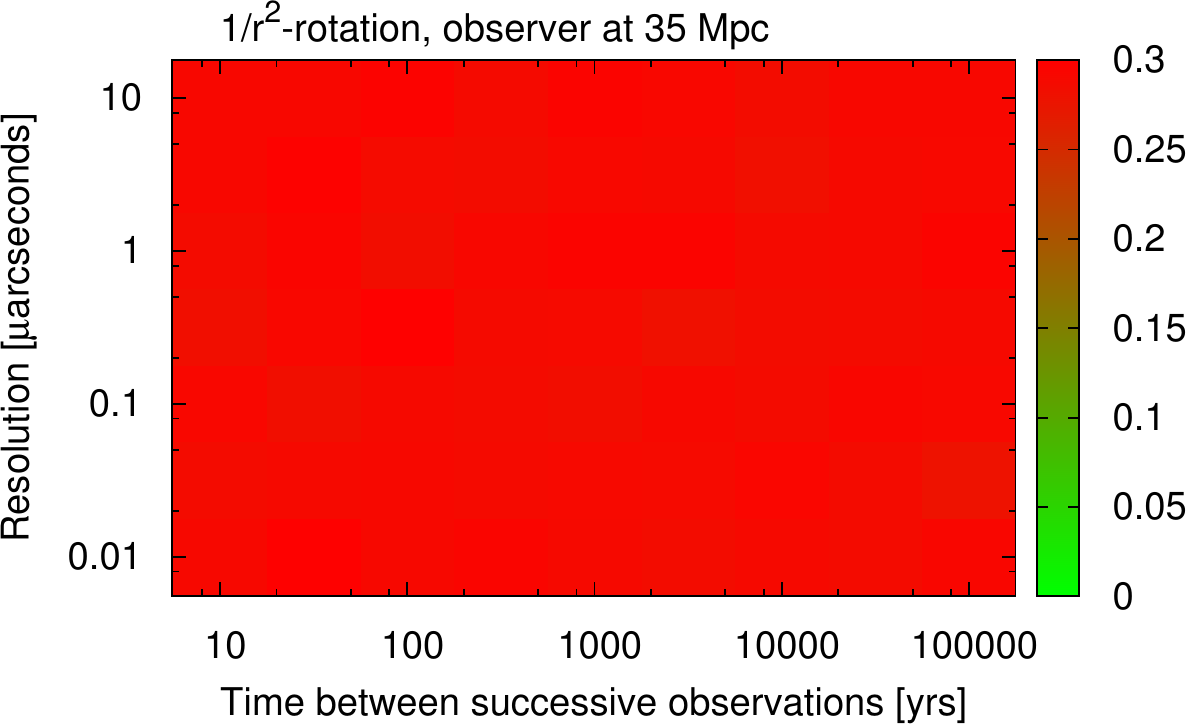}
\end{tabular} }
\end{center}
\caption{Same as figure~\ref{fig:2dplanebigsignal}, but for the inner product A, $ \hat v^{(1,1)}_{\left<F\right>(\theta,\phi)}\cdot \hat v^{(1,1)}_{\left<F^2\right>(\theta,\phi)} $.
 For this inner product, rotating and large local void models predict opposite values (0 versus 1), and are hence distinguishable from one another in principle. However, the signal  becomes significant only after 100 years of observation in the case of rotation models, and never for the large local void model.}
\label{fig:plane_detectable}
\end{figure}

For an inner product that does distinguish between rotation and a large local void we must look at figure~\ref{fig:plane_detectable}, which shows the detectability in the same way as figure~\ref{fig:2dplanebigsignal} but for the observable~A, $ \hat v^{(1,1)}_{\left<F\right>(\theta,\phi)}\cdot \hat v^{(1,1)}_{\left<F^2\right>(\theta,\phi)} $. For this inner product, the large local void scenario predicts a value of 1 while rotating models predict a value of 0. Unfortunately, here only the $1/r$-rotation and the solid rotation models are detectable after 100 years at a survey resolution of 1~$\mu$as, or, alternatively, after 10 years at a resolution of 0.1~$\mu$as.
None of the other anisotropic models produces a signal even with a thousand-fold increase in the time gap and at a much better survey resolution.  That the void model is undetectable in the inner product~A but measurable in the observable~B can be traced to the fact that the observer lives too close to the origin:  the small distance suppresses the
 $\langle F^2 \rangle$ dipole signal by a factor $(s_o/r)^3$, while the quadrupole suffers only a $(s_o/r)^2$  reduction (see discussions in section~\ref{sec:ltb}).  Taking into account peculiar velocities and detection errors, the $\langle F^2 \rangle$ dipole signal becomes completely swamped out even for  unrealistically long observation time gaps.
In contrast, an observer near the centre of a rotating region sees a  $\langle F^2 \rangle$ dipole signal suppressed only by~$s_o/r$, as shown in section~\ref{sec:origin}.  Thus, depending on the angular velocity model, rotation scenarios continue to be detectable even in the observable~A.

\subsubsection{Dependence on observer's position}

In the previous section we have for each anisotropic model placed the observer at a {\it single} location at which the local peculiar velocity marginally agrees with observational bounds from the CMB dipole.  It is now interesting to ask, how unique is such a location?  Are there other observer locations that are observationally consistent {\it and}, at the same time, conducive to detectable signals?  

Figure~\ref{fig:forreferee} shows the dependence of the detectability of products A and B on the observer position, assuming an observational gap of 10 years and for two different survey resolutions (1 and 0.01 $\mu$as). Clearly, the first and foremost factor that determines the detectability is the observer's own velocity and to an extent the velocities in the immediate vicinity.   In the case of the void model, compliance with CMB observational constraints forces the observer to remain within a rather finely-tuned 35~Mpc from the origin.  At a survey resolution of 1~$\mu$as the amount of fine-tuning is further exacerbated by the fact that the signal is rapidly lost as the observer moves towards the origin, as evidenced by the steep rise of the detectability curve.  However, if the survey resolution should improve to 0.01~$\mu$as, then at least observable B has a chance of detection at other observer positions within $35$~Mpc from the origin.

\newcolumntype{C}{ >{\centering\arraybackslash} m{0.4\textwidth} }
\newcolumntype{D}{ >{\centering\arraybackslash} m{0.2\textwidth} }

\begin{figure}
\begin{tabular}{C C D}
\hspace{.1\textwidth}B: $ \hat v^{(1,1)}_{\left<F \right>(\theta,\phi)}\cdot \hat v^{(2,2)}_{\left<F^2\right>(\theta,\phi)} $
&
A: $ \hat v^{(1,1)}_{\left<F\right>(\theta,\phi)}\cdot \hat v^{(1,1)}_{\left<F^2\right>(\theta,\phi)} $&\\
{\includegraphics[width=0.45\textwidth]{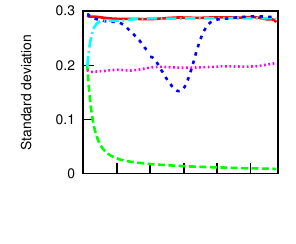}}&
\hspace{-0.15\textwidth}{\includegraphics[width=0.45\textwidth]{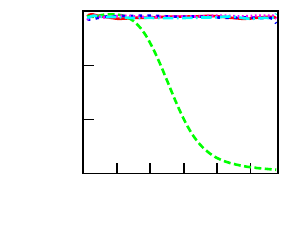}} &\vspace{-0.2\textwidth}\hspace{-0.15\textwidth}\phantom{M}\begin{tabular}{c}\includegraphics[width=0.25\textwidth]{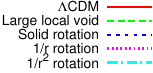}\\Resolution:\\ $1\mu$as\end{tabular}
\\
\vspace{-0.1\textwidth}\includegraphics[width=0.45\textwidth]{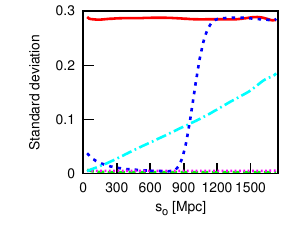}&
\vspace{-0.1\textwidth}\hspace{-0.15\textwidth}\includegraphics[width=0.45\textwidth]{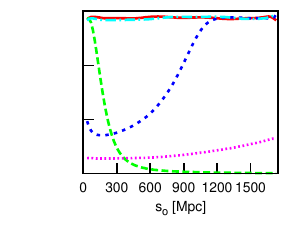}&\vspace{-0.2\textwidth}\hspace{-0.1\textwidth} \begin{tabular}{c}Resolution:\\ $0.01\mu$as\end{tabular}\\
\end{tabular}
\caption{Detectability of the inner products A (right column) and B (left column) at two fixed resolutions, 1 $\mu$as (top row) and 0.01 $\mu$as (bottom row), as a function of the observer's distance from the origin.}\label{fig:forreferee}
\end{figure}

In contrast, all rotation models considered in this work impose significantly less constraints on the observer location.
Take for example the case of $1/r$-rotation.  The relatively flat curves indicate that the detectability of both observables A and B are practically unaffected by the observer position (as long as the observer is inside the rotation region), simply because the rotation velocities in this model are independent of an object's distance from the origin.   This also implies that our results from section~\ref{sec:consistent} are not unique to the chosen observer position of $s_o =35$~Mpc from the origin, but apply to any observer position within the rotating region ($s < 2$~Gpc).  Detectability is only limited by the survey resolution.

For the solid rotation model, the rotation speed increases with distance from the origin up to $s = 1$~Gpc, beyond which it drops to zero.  This feature of the model is best exemplified by the ``dip'' in the 1-$\mu$as-detectability curve in figure~\ref{fig:forreferee} for observable B  at $s_0 \sim 800$~Mpc, which,  unsurprisingly, is also the region in which the observer's own peculiar velocity is strongly at odds with observational constraints from the CMB dipole.  
Thus, at a survey resolution of 1~$\mu$as, to maximise detectability and simultaneously maintain observational consistency the observer must sit at a distance less than but as close as possible to $s_o=690$~Mpc from the origin.  This also sets the limit of applicability of our results  in section~\ref{sec:consistent} for the solid rotation model at the said resolution.  
The limitation does not apply at a finer survey resolution of~0.01~$\mu$as, however; as shown in  figure~\ref{fig:forreferee}, at 0.01~$\mu$as all observer positions at less than $690$~Mpc from the origin yield detectable signals in both observables~A and~B.

Lastly, in the case of $1/r^2$-rotation, the observer must sit beyond $35$~Mpc from the origin in order to comply with CMB constraints.  However, at a survey resolution of 1~$\mu$as figure~\ref{fig:forreferee} shows that detectability can only be assured for observable B if the observer does not stray too far beyond $s_o = 35$~Mpc.    Improving the survey resolution to 0.01~$\mu$as allows for a positive detection even if the observer should sit as far as 2~Gpc from the origin.

\section{Discussion and conclusions}\label{sec:VI}

In this paper we have presented a new method to distinguish between isotropic cosmological models and models with large-scale anisotropies.
Our method is based on tracking the angular displacements of objects in the sky over an extended period of time, and then performing a multipole-vector decomposition of the resulting displacement maps.  While isotropic cosmologies produce no signal, different classes of anisotropic models predict unique sets of multipole vectors.
The method adds to the very limited number of tests that can be performed in order to distinguish anisotropic models from standard FLRW cosmologies and amongst themselves.

As toy models, we have considered in this work a spherically expanding LTB void and three different models with large-scale rotation. The LTB void model was a popular alternative to $\Lambda$CDM as it provided a viable explanation for the apparent acceleration of the universal expansion without invoking dark energy, albeit at the cost of forgoing the Copernican Principle. The rotation models, on the other hand, have been introduced in part as dummy alternatives in order to test if different classes of anisotropic cosmological models could be differentiated from one another using our method.  However, we note that while cosmological models with substantial global rotation are in general inconsistent with inflationary cosmologies (because of dilution during inflation), a small degree of rotational motion could conceivably  have been generated by nonlinear effects in the early universe and survived to the present time (see,  e.g.,~\cite{Christopherson:2010dw} for a discussion of vorticity evolution).
Alternatively and perhaps more realistically, extended regions of space possessing a finite angular momentum could have arisen from nonlinear structure formation at late times.
If we happened to be living in such a region, an apparent rotation of the local celestial objects is a possible manifestation of this eventuality.

Our analysis shows that, with just ten years of patience, observations of 500.000 quasars with a next-generation GAIA-like satellite at an angular resolution of 1~$\mu$as can effectively differentiate between the $\Lambda$CDM and anisotropic universes, be they LTB or rotation models. The required resolution is roughly a factor of ten better than GAIA's nominal specification.
To distinguish between different classes of anisotropic models is however a more demanding task: at a survey resolution of 1~$\mu$as, we find that a waiting time of at least 100 years is required in order for the relevant signatures to become detectable.  Furthermore, in some cases the observer location may require some degree of fine-tuning in order to maximise the signal and at the same time comply with existing cosmological bounds on the observer's peculiar velocity.  We caution at this point that we have assumed the SDSS quasar selection function to be representative for GAIA. This was a choice of convenience; a wildly different selection function could conceivably change the outcome.

Notwithstanding the somewhat discouraging conclusions, our analysis clearly demonstrates that the two classes of anisotropic universes under consideration give rise to very different signals in the real-time cosmological tests proposed in this work, signals whose detectability increases with both the observational time gap and the survey resolution.  
A possible next step to further this analysis is to include an {\it additional} peculiar velocity component for the observer, one that mimics, e.g., the peculiar motion of the solar system within the Local Group, and in a direction uncorrelated with the coherent rotation/expansion motion of the anisotropic model.  Likewise, 
relativistic effects not considered in this work such as rotation-induced frame-dragging which alters the space-time metric and subsequently the null geodesics, or relativistic aberration arising from the motion of the emitting quasars relative to the observer may constitute additional sources of uncertainties.  A detailed investigation is required to discern their full impact on our conclusions.
Furthermore, from a statistical point of view,
our analysis so far has focussed on identifying the survey parameters at which the observation of several pre-defined inner products individually transits from noise domination to signal domination.  A logical extension to this would be to consider the {\it combined} probability of the different inner products, which should  serve to enhance information extraction from the data.  Other means to abstract a stronger signal from the noise may also be possible.

In the end, survey resolution is the key.  And should observations significantly more well-resolved than GAIA become available in the future, telling apart different cosmological models---isotropic as well as anisotropic---within a reasonable timespan may not be an entirely impossible task.

\section*{Acknowledgements}
WV thanks Tristan du Pree for useful discussions.  WV is supported by a Veni research grant from the Netherlands Organisation for Scientific Research (NWO).


\end{document}